\documentclass[12pt]{article}
\usepackage{epsfig,amsfonts,amssymb}
\usepackage{hyperref}
\usepackage{cite}
\input epsf.sty
\usepackage{cite}

\def\ZZZ{{\hbox{ Z\kern-1.6mm Z}}}
\def\RRR{{\hbox{ R\kern-2.4mm R}}}
\def\CCC{{\hbox{ C\kern-2.0mm C}}}
\def\zzz{{\hbox{z\kern-1mm z}}}

\newcommand{\qeq}{{\hbox{=\kern-2.3mm ? \kern.5mm }}}
\renewcommand{\qeq}{=}

\newcommand{\eps}{\epsilon}

\newcommand{\ve}{\varepsilon}

\newcommand{\DD}{{\cal D}}

\newcommand{\wh}{\widehat}

\newcommand{\RR}{{\cal R}}

\newcommand{\be}{\begin{equation}}
\newcommand{\ee}{\end{equation}}
\newcommand{\ben}{\begin{eqnarray}\displaystyle}
\newcommand{\een}{\end{eqnarray}}

\newcommand{\refb}[1]{(\ref{#1})}

\newcommand{\sectiono}[1]{\section{#1}\setcounter{equation}{0}}

\def\one{{\hbox{ 1\kern-.8mm l}}}
\def\zero{{\hbox{ 0\kern-1.5mm 0}}}

\newcommand{\bea}[1]{\begin{eqnarray}\label{#1} }
\newcommand{\eea}{\end{eqnarray}}

\newcommand{\eqref}{\refb}

\def\figbox{

\def\JPicScale{0.8}
\ifx\JPicScale\undefined\def\JPicScale{1}\fi
\unitlength \JPicScale mm
\begin{picture}(90,70)(0,0)
\linethickness{0.3mm}
\put(40,60){\line(1,0){40}}
\linethickness{0.3mm}
\put(80,20){\line(0,1){40}}
\linethickness{0.3mm}
\put(40,20){\line(1,0){40}}
\linethickness{0.3mm}
\put(40,20){\line(0,1){40}}
\linethickness{0.3mm}
\put(20,20){\line(1,0){20}}
\linethickness{0.3mm}
\linethickness{0.3mm}
\put(40,0){\line(0,1){20}}
\linethickness{0.3mm}
\put(20,60){\line(1,0){20}}
\linethickness{0.3mm}
\put(40,60){\line(0,1){20}}
\linethickness{0.3mm}
\put(80,60){\line(0,1){20}}
\linethickness{0.3mm}
\put(80,60){\line(1,0){20}}
\linethickness{0.3mm}
\put(80,0){\line(0,1){20}}
\linethickness{0.3mm}
\put(80,20){\line(1,0){20}}
\put(35,15){\makebox(0,0)[cc]{$\nearrow$}}

\put(35,65){\makebox(0,0)[cc]{$\searrow$}}

\put(85,15){\makebox(0,0)[cc]{$\nwarrow$}}

\put(85,65){\makebox(0,0)[cc]{$\swarrow$}}

\put(60,15){\makebox(0,0)[cc]{$\ell\rightarrow$}}

\put(52,40){\makebox(0,0)[cc]{$\uparrow (p_A-\ell)$}}

\put(60,65){\makebox(0,0)[cc]{$(p_A+p_B-\ell)\rightarrow$}}

\put(92,40){\makebox(0,0)[cc]{$\uparrow (\ell+p_C)$}}

\put(30,10){\makebox(0,0)[cc]{$p_A$}}

\put(30,70){\makebox(0,0)[cc]{$p_B$}}

\put(90,10){\makebox(0,0)[cc]{$p_C$}}

\put(90,70){\makebox(0,0)[cc]{$p_D$}}

\end{picture}

}

\def\figboxtwo{

\def\JPicScale{0.6}
\ifx\JPicScale\undefined\def\JPicScale{1}\fi
\unitlength \JPicScale mm
\begin{picture}(130,70)(0,0)
\linethickness{0.3mm}
\multiput(30,50)(0.3,0.12){167}{\line(1,0){0.3}}
\linethickness{0.3mm}
\multiput(80,70)(0.3,-0.12){167}{\line(1,0){0.3}}
\linethickness{0.3mm}
\multiput(80,30)(0.3,0.12){167}{\line(1,0){0.3}}
\linethickness{0.3mm}
\multiput(30,50)(0.3,-0.12){167}{\line(1,0){0.3}}
\put(50,35){\makebox(0,0)[cc]{1}}

\put(105,35){\makebox(0,0)[cc]{2}}

\put(50,65){\makebox(0,0)[cc]{3}}

\put(105,65){\makebox(0,0)[cc]{4}}

\put(55,45){\makebox(0,0)[cc]{$\ell$}}

\put(125,40){\makebox(0,0)[cc]{$\ell+p_C$}}

\put(60,55){\makebox(0,0)[cc]{$p_A-\ell$}}

\put(135,60){\makebox(0,0)[cc]{$p_A+p_B-\ell$}}

\end{picture}

}

\def\figAone{

\def\JPicScale{0.6}
\ifx\JPicScale\undefined\def\JPicScale{1}\fi
\unitlength \JPicScale mm
\begin{picture}(120,60)(0,0)
\linethickness{0.3mm}
\multiput(10,40)(0.12,0.12){167}{\line(1,0){0.12}}
\linethickness{0.3mm}
\multiput(30,60)(0.12,-0.12){167}{\line(1,0){0.12}}
\linethickness{0.3mm}
\multiput(30,20)(0.12,0.12){167}{\line(1,0){0.12}}
\linethickness{0.3mm}
\multiput(10,40)(0.12,-0.12){167}{\line(1,0){0.12}}
\linethickness{.6mm}
\put(20,25){\line(0,1){10}}
\put(15,30){\makebox(0,0)[cc]{1}}

\put(45,30){\makebox(0,0)[cc]{2}}

\put(15,50){\makebox(0,0)[cc]{3}}

\put(45,50){\makebox(0,0)[cc]{4}}

\put(65,40){\makebox(0,0)[cc]{=}}

\linethickness{.3mm}
\multiput(80,40)(0.12,0.12){167}{\line(1,0){0.12}}
\linethickness{0.3mm}
\multiput(100,60)(0.12,-0.12){167}{\line(1,0){0.12}}
\linethickness{0.3mm}
\multiput(100,20)(0.12,0.12){167}{\line(1,0){0.12}}
\linethickness{0.3mm}
\put(90,20){\line(1,0){10}}
\linethickness{0.3mm}
\put(80,40){\line(1,0){10}}
\put(95,15){\makebox(0,0)[cc]{1}}

\put(85,35){\makebox(0,0)[cc]{1}}

\put(115,30){\makebox(0,0)[cc]{2}}

\put(85,50){\makebox(0,0)[cc]{3}}

\put(115,50){\makebox(0,0)[cc]{4}}

\end{picture}

}

\def\figAtwo{

\def\JPicScale{0.6}
\ifx\JPicScale\undefined\def\JPicScale{1}\fi
\unitlength \JPicScale mm
\begin{picture}(120,60)(0,0)
\linethickness{0.3mm}
\multiput(10,40)(0.12,0.12){167}{\line(1,0){0.12}}
\linethickness{0.3mm}
\multiput(30,60)(0.12,-0.12){167}{\line(1,0){0.12}}
\linethickness{0.3mm}
\multiput(30,20)(0.12,0.12){167}{\line(1,0){0.12}}
\linethickness{0.3mm}
\multiput(10,40)(0.12,-0.12){167}{\line(1,0){0.12}}
\linethickness{.6mm}
\put(40,25){\line(0,1){10}}
\put(15,30){\makebox(0,0)[cc]{1}}

\put(45,30){\makebox(0,0)[cc]{2}}

\put(15,50){\makebox(0,0)[cc]{3}}

\put(45,50){\makebox(0,0)[cc]{4}}

\put(65,40){\makebox(0,0)[cc]{=}}

\linethickness{0.3mm}
\multiput(80,40)(0.12,0.12){167}{\line(1,0){0.12}}
\linethickness{0.3mm}
\multiput(100,60)(0.12,-0.12){167}{\line(1,0){0.12}}
\linethickness{0.3mm}
\put(100,20){\line(1,0){10}}
\linethickness{0.3mm}
\put(110,40){\line(1,0){10}}
\put(85,30){\makebox(0,0)[cc]{1}}

\put(115,35){\makebox(0,0)[cc]{2}}

\put(85,50){\makebox(0,0)[cc]{3}}

\put(115,50){\makebox(0,0)[cc]{4}}

\linethickness{0.3mm}
\multiput(80,40)(0.12,-0.12){167}{\line(1,0){0.12}}
\put(105,15){\makebox(0,0)[cc]{2}}

\end{picture}

}

\def\figAthree{

\def\JPicScale{0.6}
\ifx\JPicScale\undefined\def\JPicScale{1}\fi
\unitlength \JPicScale mm
\begin{picture}(120,60)(0,0)
\linethickness{0.3mm}
\multiput(10,40)(0.12,0.12){167}{\line(1,0){0.12}}
\linethickness{0.3mm}
\multiput(30,60)(0.12,-0.12){167}{\line(1,0){0.12}}
\linethickness{0.3mm}
\multiput(30,20)(0.12,0.12){167}{\line(1,0){0.12}}
\linethickness{0.3mm}
\multiput(10,40)(0.12,-0.12){167}{\line(1,0){0.12}}
\linethickness{.6mm}
\put(40,25){\line(0,1){10}}
\put(15,30){\makebox(0,0)[cc]{1}}

\put(45,30){\makebox(0,0)[cc]{2}}

\put(15,50){\makebox(0,0)[cc]{3}}

\put(45,50){\makebox(0,0)[cc]{4}}

\put(65,40){\makebox(0,0)[cc]{=}}

\linethickness{0.3mm}
\multiput(80,40)(0.12,0.12){167}{\line(1,0){0.12}}
\linethickness{0.3mm}
\multiput(100,60)(0.12,-0.12){167}{\line(1,0){0.12}}
\linethickness{0.3mm}
\put(100,20){\line(1,0){10}}
\linethickness{0.3mm}
\put(110,40){\line(1,0){10}}
\put(85,35){\makebox(0,0)[cc]{1}}

\put(115,35){\makebox(0,0)[cc]{2}}

\put(85,50){\makebox(0,0)[cc]{3}}

\put(115,50){\makebox(0,0)[cc]{4}}

\put(105,15){\makebox(0,0)[cc]{2}}

\linethickness{.6mm}
\put(20,25){\line(0,1){10}}
\linethickness{0.3mm}
\put(80,40){\line(1,0){10}}
\linethickness{0.3mm}
\multiput(90,30)(0.12,-0.12){83}{\line(1,0){0.12}}
\put(90,25){\makebox(0,0)[cc]{1}}

\end{picture}

}

\def\figAonea{

\def\JPicScale{0.6}
\ifx\JPicScale\undefined\def\JPicScale{1}\fi
\unitlength \JPicScale mm
\begin{picture}(110,80)(0,0)
\linethickness{0.3mm}
\multiput(10,50)(0.12,0.12){167}{\line(1,0){0.12}}
\linethickness{0.3mm}
\multiput(30,70)(0.12,-0.12){167}{\line(1,0){0.12}}
\linethickness{0.3mm}
\multiput(30,30)(0.12,0.12){167}{\line(1,0){0.12}}
\linethickness{0.3mm}
\multiput(10,50)(0.12,-0.12){167}{\line(1,0){0.12}}
\linethickness{.6mm}
\put(20,35){\line(0,1){10}}
\linethickness{.6mm}
\put(25,30){\line(0,1){40}}
\linethickness{0.3mm}
\multiput(70,50)(0.12,0.12){167}{\line(1,0){0.12}}
\linethickness{0.3mm}
\multiput(90,70)(0.12,-0.12){167}{\line(1,0){0.12}}
\linethickness{0.3mm}
\multiput(90,30)(0.12,0.12){167}{\line(1,0){0.12}}
\linethickness{0.3mm}
\multiput(70,50)(0.12,-0.12){167}{\line(1,0){0.12}}
\linethickness{.6mm}
\put(80,35){\line(0,1){10}}
\linethickness{.6mm}
\qbezier(100,80)(100.03,74.84)(97.62,67.62)
\qbezier(97.62,67.62)(95.22,60.41)(90,50)
\qbezier(90,50)(84.78,39.58)(82.38,33.56)
\qbezier(82.38,33.56)(79.97,27.55)(80,25)
\qbezier(80,25)(80,22.39)(80,21.19)
\qbezier(80,21.19)(80,19.98)(80,20)
\put(60,50){\makebox(0,0)[cc]{+}}

\put(120,50){\makebox(0,0)[cc]{=}}

\end{picture}

}

\def\figAoneb{

\def\JPicScale{0.6}
\ifx\JPicScale\undefined\def\JPicScale{1}\fi
\unitlength \JPicScale mm
\begin{picture}(110,80)(0,0)
\linethickness{0.3mm}
\multiput(10,50)(0.12,0.12){167}{\line(1,0){0.12}}
\linethickness{0.3mm}
\multiput(30,70)(0.12,-0.12){167}{\line(1,0){0.12}}
\linethickness{0.3mm}
\linethickness{0.3mm}
\multiput(30,30)(0.12,0.12){167}{\line(1,0){0.12}}
\linethickness{0.3mm}
\put(10,50){\line(1,0){20}}
\linethickness{0.3mm}
\put(10,30){\line(1,0){20}}
\linethickness{.6mm}
\put(20,20){\line(0,1){50}}
\linethickness{0.3mm}
\multiput(70,50)(0.12,0.12){167}{\line(1,0){0.12}}
\linethickness{0.3mm}
\multiput(90,70)(0.12,-0.12){167}{\line(1,0){0.12}}
\linethickness{0.3mm}
\multiput(90,30)(0.12,0.12){167}{\line(1,0){0.12}}
\linethickness{0.3mm}
\put(70,30){\line(1,0){20}}
\linethickness{0.3mm}
\put(70,50){\line(1,0){20}}
\linethickness{.6mm}
\qbezier(100,80)(97.41,69.59)(94.41,62.97)
\qbezier(94.41,62.97)(91.4,56.35)(87.5,52.5)
\qbezier(87.5,52.5)(83.59,48.66)(81.78,40.84)
\qbezier(81.78,40.84)(79.98,33.02)(80,20)
\put(60,50){\makebox(0,0)[cc]{+}}

\end{picture}

}

\def\figAtwoa{

\def\JPicScale{0.6}
\ifx\JPicScale\undefined\def\JPicScale{1}\fi
\unitlength \JPicScale mm
\begin{picture}(110,80)(0,0)
\linethickness{0.3mm}
\multiput(10,50)(0.12,0.12){167}{\line(1,0){0.12}}
\linethickness{0.3mm}
\multiput(30,70)(0.12,-0.12){167}{\line(1,0){0.12}}
\linethickness{0.3mm}
\multiput(30,30)(0.12,0.12){167}{\line(1,0){0.12}}
\linethickness{0.3mm}
\multiput(10,50)(0.12,-0.12){167}{\line(1,0){0.12}}
\linethickness{.6mm}
\put(40,35){\line(0,1){10}}
\linethickness{.6mm}
\put(20,30){\line(0,1){40}}
\linethickness{0.3mm}
\multiput(70,50)(0.12,0.12){167}{\line(1,0){0.12}}
\linethickness{0.3mm}
\multiput(90,70)(0.12,-0.12){167}{\line(1,0){0.12}}
\linethickness{0.3mm}
\multiput(90,30)(0.12,0.12){167}{\line(1,0){0.12}}
\linethickness{0.3mm}
\multiput(70,50)(0.12,-0.12){167}{\line(1,0){0.12}}
\linethickness{.6mm}
\put(100,35){\line(0,1){10}}
\linethickness{.6mm}
\qbezier(100,80)(100.03,74.84)(97.62,67.62)
\qbezier(97.62,67.62)(95.22,60.41)(90,50)
\qbezier(90,50)(84.78,39.58)(82.38,33.56)
\qbezier(82.38,33.56)(79.97,27.55)(80,25)
\qbezier(80,25)(80,22.39)(80,21.19)
\qbezier(80,21.19)(80,19.98)(80,20)
\put(60,50){\makebox(0,0)[cc]{+}}

\put(120,50){\makebox(0,0)[cc]{=}}

\end{picture}

}

\def\figAtwob{

\ifx\JPicScale\undefined\def\JPicScale{1}\fi
\unitlength \JPicScale mm
\begin{picture}(110,80)(0,0)
\linethickness{0.3mm}
\multiput(10,50)(0.12,0.12){167}{\line(1,0){0.12}}
\linethickness{0.3mm}
\multiput(30,70)(0.12,-0.12){167}{\line(1,0){0.12}}
\linethickness{0.3mm}
\multiput(10,50)(0.12,-0.12){167}{\line(1,0){0.12}}
\linethickness{0.3mm}
\put(30,30){\line(1,0){20}}
\linethickness{0.3mm}
\put(30,50){\line(1,0){20}}
\linethickness{.6mm}
\put(20,30){\line(0,1){40}}
\linethickness{0.3mm}
\multiput(70,50)(0.12,0.12){167}{\line(1,0){0.12}}
\linethickness{0.3mm}
\multiput(90,70)(0.12,-0.12){167}{\line(1,0){0.12}}
\linethickness{0.3mm}
\put(100,50){\line(1,0){10}}
\linethickness{0.3mm}
\multiput(70,50)(0.12,-0.12){167}{\line(1,0){0.12}}
\linethickness{0.3mm}
\put(90,30){\line(1,0){20}}
\linethickness{.6mm}
\qbezier(105,80)(105.02,72.2)(103.22,66.19)
\qbezier(103.22,66.19)(101.41,60.17)(97.5,55)
\qbezier(97.5,55)(93.6,49.8)(90.59,45.59)
\qbezier(90.59,45.59)(87.59,41.38)(85,37.5)
\qbezier(85,37.5)(82.39,33.62)(81.19,29.41)
\qbezier(81.19,29.41)(79.98,25.2)(80,20)
\put(60,50){\makebox(0,0)[cc]{+}}

\end{picture}

}

\def\figAonec{

\def\JPicScale{0.6}
\ifx\JPicScale\undefined\def\JPicScale{1}\fi
\unitlength \JPicScale mm
\begin{picture}(110,80)(0,0)
\linethickness{0.3mm}
\multiput(10,50)(0.12,0.12){167}{\line(1,0){0.12}}
\linethickness{0.3mm}
\multiput(30,70)(0.12,-0.12){167}{\line(1,0){0.12}}
\linethickness{0.3mm}
\multiput(30,30)(0.12,0.12){167}{\line(1,0){0.12}}
\linethickness{0.3mm}
\multiput(10,50)(0.12,-0.12){167}{\line(1,0){0.12}}
\linethickness{.6mm}%
\put(20,35){\line(0,1){10}}
\linethickness{.6mm}%
\put(40,30){\line(0,1){40}}
\linethickness{0.3mm}
\multiput(70,50)(0.12,0.12){167}{\line(1,0){0.12}}
\linethickness{0.3mm}
\multiput(90,70)(0.12,-0.12){167}{\line(1,0){0.12}}
\linethickness{0.3mm}
\multiput(90,30)(0.12,0.12){167}{\line(1,0){0.12}}
\linethickness{0.3mm}
\multiput(70,50)(0.12,-0.12){167}{\line(1,0){0.12}}
\linethickness{.6mm}%
\put(80,35){\line(0,1){10}}
\put(60,50){\makebox(0,0)[cc]{+}}

\put(120,50){\makebox(0,0)[cc]{=}}

\linethickness{.6mm}%
\qbezier(75,80)(74.98,74.81)(76.78,70)
\qbezier(76.78,70)(78.59,65.19)(82.5,60)
\qbezier(82.5,60)(86.39,54.8)(90,50.59)
\qbezier(90,50.59)(93.61,46.38)(97.5,42.5)
\qbezier(97.5,42.5)(101.41,38.63)(103.22,33.22)
\qbezier(103.22,33.22)(105.02,27.8)(105,20)
\end{picture}

}

\def\figAoned{

\def\JPicScale{0.6}
\ifx\JPicScale\undefined\def\JPicScale{1}\fi
\unitlength \JPicScale mm
\begin{picture}(110,80)(0,0)
\linethickness{0.3mm}
\multiput(10,50)(0.12,0.12){167}{\line(1,0){0.12}}
\linethickness{0.3mm}
\multiput(30,70)(0.12,-0.12){167}{\line(1,0){0.12}}
\linethickness{0.3mm}
\multiput(30,30)(0.12,0.12){167}{\line(1,0){0.12}}
\linethickness{.6mm}%
\put(40,30){\line(0,1){40}}
\linethickness{0.3mm}
\multiput(70,50)(0.12,0.12){167}{\line(1,0){0.12}}
\linethickness{0.3mm}
\multiput(90,70)(0.12,-0.12){167}{\line(1,0){0.12}}
\linethickness{0.3mm}
\multiput(90,30)(0.12,0.12){167}{\line(1,0){0.12}}
\put(60,50){\makebox(0,0)[cc]{+}}

\linethickness{.6mm}%
\qbezier(70,80)(69.98,74.81)(71.78,70)
\qbezier(71.78,70)(73.59,65.19)(77.5,60)
\qbezier(77.5,60)(81.39,54.8)(85,50.59)
\qbezier(85,50.59)(88.61,46.38)(92.5,42.5)
\qbezier(92.5,42.5)(96.41,38.63)(98.22,33.22)
\qbezier(98.22,33.22)(100.02,27.8)(100,20)
\linethickness{0.3mm}
\put(10,50){\line(1,0){20}}
\linethickness{0.3mm}
\put(10,30){\line(1,0){20}}
\linethickness{0.3mm}
\put(70,50){\line(1,0){10}}
\linethickness{0.3mm}
\put(70,30){\line(1,0){20}}
\linethickness{0.3mm}
\end{picture}

}

\def\figAtwoc{

\def\JPicScale{0.6}
\ifx\JPicScale\undefined\def\JPicScale{1}\fi
\unitlength \JPicScale mm
\begin{picture}(110,80)(0,0)
\linethickness{0.3mm}
\multiput(10,50)(0.12,0.12){167}{\line(1,0){0.12}}
\linethickness{0.3mm}
\multiput(30,70)(0.12,-0.12){167}{\line(1,0){0.12}}
\linethickness{0.3mm}
\multiput(10,50)(0.12,-0.12){167}{\line(1,0){0.12}}
\linethickness{.6mm}%
\put(35,25){\line(0,1){50}}
\linethickness{0.3mm}
\multiput(70,50)(0.12,0.12){167}{\line(1,0){0.12}}
\linethickness{0.3mm}
\multiput(90,70)(0.12,-0.12){167}{\line(1,0){0.12}}
\linethickness{0.3mm}
\multiput(70,50)(0.12,-0.12){167}{\line(1,0){0.12}}
\put(60,50){\makebox(0,0)[cc]{+}}

\put(120,50){\makebox(0,0)[cc]{=}}

\linethickness{.6mm}
\qbezier(75,80)(74.98,74.8)(76.78,70.59)
\qbezier(76.78,70.59)(78.59,66.38)(82.5,62.5)
\qbezier(82.5,62.5)(86.41,58.62)(88.81,54.41)
\qbezier(88.81,54.41)(91.22,50.2)(92.5,45)
\qbezier(92.5,45)(93.8,39.83)(94.41,33.81)
\qbezier(94.41,33.81)(95.01,27.8)(95,20)
\linethickness{0.3mm}
\multiput(30,30)(0.12,0.12){167}{\line(1,0){0.12}}
\linethickness{0.3mm}
\multiput(90,30)(0.12,0.12){167}{\line(1,0){0.12}}
\linethickness{.6mm}
\put(40,35){\line(0,1){10}}
\linethickness{.6mm}
\put(100,35){\line(0,1){10}}
\end{picture}

}

\def\figAtwod{

\def\JPicScale{0.6}
\ifx\JPicScale\undefined\def\JPicScale{1}\fi
\unitlength \JPicScale mm
\begin{picture}(110,80)(0,0)
\linethickness{0.3mm}
\multiput(10,50)(0.12,0.12){167}{\line(1,0){0.12}}
\linethickness{0.3mm}
\multiput(30,70)(0.12,-0.12){167}{\line(1,0){0.12}}
\linethickness{0.3mm}
\multiput(10,50)(0.12,-0.12){167}{\line(1,0){0.12}}
\linethickness{0.3mm}
\put(30,30){\line(1,0){20}}
\linethickness{0.3mm}
\put(30,50){\line(1,0){20}}
\linethickness{.6mm}
\put(40,20){\line(0,1){50}}
\linethickness{0.3mm}
\multiput(70,50)(0.12,0.12){167}{\line(1,0){0.12}}
\linethickness{0.3mm}
\multiput(90,70)(0.12,-0.12){167}{\line(1,0){0.12}}
\linethickness{0.3mm}
\put(90,50){\line(1,0){20}}
\linethickness{0.3mm}
\multiput(70,50)(0.12,-0.12){167}{\line(1,0){0.12}}
\linethickness{0.3mm}
\put(90,30){\line(1,0){20}}
\put(60,50){\makebox(0,0)[cc]{+}}

\linethickness{.6mm}
\qbezier(80,80)(79.98,74.8)(81.78,70.59)
\qbezier(81.78,70.59)(83.59,66.38)(87.5,62.5)
\qbezier(87.5,62.5)(91.41,58.62)(93.81,54.41)
\qbezier(93.81,54.41)(96.22,50.2)(97.5,45)
\qbezier(97.5,45)(98.8,39.83)(99.41,33.81)
\qbezier(99.41,33.81)(100.01,27.8)(100,20)
\end{picture}

}

\def\figAonetwob{

\def\JPicScale{0.6}
\ifx\JPicScale\undefined\def\JPicScale{1}\fi
\unitlength \JPicScale mm
\begin{picture}(110,80)(0,0)
\linethickness{0.3mm}
\multiput(10,50)(0.12,0.12){167}{\line(1,0){0.12}}
\linethickness{0.3mm}
\multiput(30,70)(0.12,-0.12){167}{\line(1,0){0.12}}
\linethickness{.6mm}
\put(25,25){\line(0,1){50}}
\linethickness{0.3mm}
\multiput(70,50)(0.12,0.12){167}{\line(1,0){0.12}}
\linethickness{0.3mm}
\multiput(90,70)(0.12,-0.12){167}{\line(1,0){0.12}}
\put(60,50){\makebox(0,0)[cc]{+}}

\linethickness{0.3mm}
\multiput(10,40)(0.24,-0.12){83}{\line(1,0){0.24}}
\linethickness{0.3mm}
\put(30,30){\line(1,0){10}}
\linethickness{0.3mm}
\multiput(10,50)(0.24,-0.12){83}{\line(1,0){0.24}}
\linethickness{0.3mm}
\put(30,50){\line(1,0){20}}
\linethickness{0.3mm}
\put(90,30){\line(1,0){20}}
\linethickness{0.3mm}
\put(100,50){\line(1,0){10}}
\linethickness{0.3mm}
\multiput(70,50)(0.24,-0.12){83}{\line(1,0){0.24}}
\linethickness{0.3mm}
\multiput(70,40)(0.24,-0.12){83}{\line(1,0){0.24}}
\linethickness{.6mm}
\qbezier(105,80)(105.01,72.19)(104.41,67.38)
\qbezier(104.41,67.38)(103.8,62.56)(102.5,60)
\qbezier(102.5,60)(101.21,57.41)(99.41,55)
\qbezier(99.41,55)(97.6,52.59)(95,50)
\qbezier(95,50)(92.41,47.4)(90,45.59)
\qbezier(90,45.59)(87.59,43.79)(85,42.5)
\qbezier(85,42.5)(82.39,41.26)(81.19,35.84)
\qbezier(81.19,35.84)(79.98,30.43)(80,20)
\end{picture}

}

\def\figAonetwoa{

\def\JPicScale{0.6}
\ifx\JPicScale\undefined\def\JPicScale{1}\fi
\unitlength \JPicScale mm
\begin{picture}(110,80)(0,0)
\linethickness{0.3mm}
\multiput(10,50)(0.12,0.12){167}{\line(1,0){0.12}}
\linethickness{0.3mm}
\multiput(30,70)(0.12,-0.12){167}{\line(1,0){0.12}}
\linethickness{0.3mm}
\multiput(10,50)(0.12,-0.12){167}{\line(1,0){0.12}}
\linethickness{.6mm}
\put(25,25){\line(0,1){50}}
\linethickness{0.3mm}
\multiput(70,50)(0.12,0.12){167}{\line(1,0){0.12}}
\linethickness{0.3mm}
\multiput(90,70)(0.12,-0.12){167}{\line(1,0){0.12}}
\linethickness{0.3mm}
\multiput(70,50)(0.12,-0.12){167}{\line(1,0){0.12}}
\put(60,50){\makebox(0,0)[cc]{+}}

\put(120,50){\makebox(0,0)[cc]{=}}

\linethickness{0.3mm}
\multiput(30,30)(0.12,0.12){167}{\line(1,0){0.12}}
\linethickness{0.3mm}
\multiput(90,30)(0.12,0.12){167}{\line(1,0){0.12}}
\linethickness{.6mm}
\put(40,35){\line(0,1){10}}
\linethickness{.6mm}
\put(100,35){\line(0,1){10}}
\linethickness{.6mm}
\put(20,35){\line(0,1){10}}
\linethickness{.6mm}
\put(80,35){\line(0,1){10}}
\linethickness{.6mm}
\qbezier(100,80)(100.02,74.81)(98.81,70)
\qbezier(98.81,70)(97.61,65.19)(95,60)
\qbezier(95,60)(92.4,54.79)(90.59,51.78)
\qbezier(90.59,51.78)(88.79,48.77)(87.5,47.5)
\qbezier(87.5,47.5)(86.2,46.27)(85.59,39.66)
\qbezier(85.59,39.66)(84.99,33.04)(85,20)
\end{picture}

}

\def\figAonetwoc{

\def\JPicScale{0.6}
\ifx\JPicScale\undefined\def\JPicScale{1}\fi
\unitlength \JPicScale mm
\begin{picture}(110,80)(0,0)
\linethickness{0.3mm}
\multiput(10,50)(0.12,0.12){167}{\line(1,0){0.12}}
\linethickness{0.3mm}
\multiput(30,70)(0.12,-0.12){167}{\line(1,0){0.12}}
\linethickness{0.3mm}
\multiput(10,50)(0.12,-0.12){167}{\line(1,0){0.12}}
\linethickness{.6mm}
\put(35,25){\line(0,1){50}}
\linethickness{0.3mm}
\multiput(70,50)(0.12,0.12){167}{\line(1,0){0.12}}
\linethickness{0.3mm}
\multiput(90,70)(0.12,-0.12){167}{\line(1,0){0.12}}
\linethickness{0.3mm}
\multiput(70,50)(0.12,-0.12){167}{\line(1,0){0.12}}
\put(60,50){\makebox(0,0)[cc]{+}}

\put(120,50){\makebox(0,0)[cc]{=}}

\linethickness{.6mm}
\qbezier(75,80)(74.98,74.8)(76.78,70.59)
\qbezier(76.78,70.59)(78.59,66.38)(82.5,62.5)
\qbezier(82.5,62.5)(86.41,58.62)(88.81,54.41)
\qbezier(88.81,54.41)(91.22,50.2)(92.5,45)
\qbezier(92.5,45)(93.8,39.83)(94.41,33.81)
\qbezier(94.41,33.81)(95.01,27.8)(95,20)
\linethickness{0.3mm}
\multiput(30,30)(0.12,0.12){167}{\line(1,0){0.12}}
\linethickness{0.3mm}
\multiput(90,30)(0.12,0.12){167}{\line(1,0){0.12}}
\linethickness{.6mm}
\put(40,35){\line(0,1){10}}
\linethickness{.6mm}
\put(100,35){\line(0,1){10}}
\linethickness{.6mm}
\put(20,35){\line(0,1){10}}
\linethickness{.6mm}
\put(80,35){\line(0,1){10}}
\end{picture}

}

\def\figAonetwod{

\def\JPicScale{0.6}
\ifx\JPicScale\undefined\def\JPicScale{1}\fi
\unitlength \JPicScale mm
\begin{picture}(110,80)(0,0)
\linethickness{0.3mm}
\multiput(10,50)(0.12,0.12){167}{\line(1,0){0.12}}
\linethickness{0.3mm}
\multiput(30,70)(0.12,-0.12){167}{\line(1,0){0.12}}
\linethickness{.6mm}
\put(35,25){\line(0,1){50}}
\linethickness{0.3mm}
\multiput(70,50)(0.12,0.12){167}{\line(1,0){0.12}}
\linethickness{0.3mm}
\multiput(90,70)(0.12,-0.12){167}{\line(1,0){0.12}}
\put(60,50){\makebox(0,0)[cc]{+}}

\linethickness{0.3mm}
\multiput(10,40)(0.24,-0.12){83}{\line(1,0){0.24}}
\linethickness{0.3mm}
\put(30,30){\line(1,0){10}}
\linethickness{0.3mm}
\multiput(10,50)(0.24,-0.12){83}{\line(1,0){0.24}}
\linethickness{0.3mm}
\put(20,50){\line(1,0){30}}
\linethickness{0.3mm}
\put(90,30){\line(1,0){20}}
\linethickness{0.3mm}
\put(90,50){\line(1,0){20}}
\linethickness{0.3mm}
\multiput(70,50)(0.24,-0.12){83}{\line(1,0){0.24}}
\linethickness{0.3mm}
\multiput(70,40)(0.24,-0.12){83}{\line(1,0){0.24}}
\linethickness{.6mm}
\qbezier(80,80)(79.98,74.8)(81.19,71.19)
\qbezier(81.19,71.19)(82.39,67.58)(85,65)
\qbezier(85,65)(87.59,62.41)(90,60)
\qbezier(90,60)(92.41,57.59)(95,55)
\qbezier(95,55)(97.6,52.41)(99.41,49.41)
\qbezier(99.41,49.41)(101.21,46.4)(102.5,42.5)
\qbezier(102.5,42.5)(103.8,38.63)(104.41,33.22)
\qbezier(104.41,33.22)(105.01,27.8)(105,20)
\end{picture}

}

\def\figAonee{

\def\JPicScale{0.6}
\ifx\JPicScale\undefined\def\JPicScale{1}\fi
\unitlength \JPicScale mm
\begin{picture}(110,70)(0,0)
\linethickness{0.3mm}
\multiput(10,40)(0.12,0.12){167}{\line(1,0){0.12}}
\linethickness{0.3mm}
\multiput(30,60)(0.12,-0.12){167}{\line(1,0){0.12}}
\linethickness{0.3mm}
\multiput(30,20)(0.12,0.12){167}{\line(1,0){0.12}}
\linethickness{0.3mm}
\multiput(10,40)(0.12,-0.12){167}{\line(1,0){0.12}}
\linethickness{.6mm}
\put(20,25){\line(0,1){10}}
\linethickness{.6mm}
\qbezier(40,10)(40.01,15.2)(39.41,19.41)
\qbezier(39.41,19.41)(38.8,23.62)(37.5,27.5)
\qbezier(37.5,27.5)(36.2,31.41)(35,33.22)
\qbezier(35,33.22)(33.8,35.02)(32.5,35)
\qbezier(32.5,35)(31.2,35.02)(30,33.81)
\qbezier(30,33.81)(28.8,32.61)(27.5,30)
\qbezier(27.5,30)(26.2,27.41)(25,25)
\qbezier(25,25)(23.8,22.59)(22.5,20)
\qbezier(22.5,20)(21.22,17.38)(18.81,16.78)
\qbezier(18.81,16.78)(16.41,16.18)(12.5,17.5)
\qbezier(12.5,17.5)(8.59,18.77)(6.78,21.78)
\qbezier(6.78,21.78)(4.98,24.79)(5,30)
\qbezier(5,30)(5,35.22)(5,37.62)
\qbezier(5,37.62)(5,40.03)(5,40)
\qbezier(5,40)(5,39.97)(5,42.38)
\qbezier(5,42.38)(5,44.78)(5,50)
\qbezier(5,50)(5,55.19)(5,60)
\qbezier(5,60)(5,64.81)(5,70)
\put(60,40){\makebox(0,0)[cc]{=}}

\linethickness{0.3mm}
\multiput(70,40)(0.12,0.12){167}{\line(1,0){0.12}}
\linethickness{0.3mm}
\multiput(90,60)(0.12,-0.12){167}{\line(1,0){0.12}}
\linethickness{0.3mm}
\multiput(90,20)(0.12,0.12){167}{\line(1,0){0.12}}
\linethickness{0.3mm}
\put(80,20){\line(1,0){10}}
\linethickness{0.3mm}
\put(70,40){\line(1,0){10}}

\linethickness{.6mm}
\qbezier(105,10)(105.02,15.19)(103.81,20)
\qbezier(103.81,20)(102.61,24.81)(100,30)
\qbezier(100,30)(97.41,35.23)(94.41,36.44)
\qbezier(94.41,36.44)(91.4,37.64)(87.5,35)
\qbezier(87.5,35)(83.6,32.39)(80.59,31.19)
\qbezier(80.59,31.19)(77.59,29.98)(75,30)
\qbezier(75,30)(72.4,29.98)(70.59,31.19)
\qbezier(70.59,31.19)(68.79,32.39)(67.5,35)
\qbezier(67.5,35)(66.2,37.55)(65.59,43.56)
\qbezier(65.59,43.56)(64.99,49.58)(65,60)

\end{picture}

}

\def\figAtwoe{

\def\JPicScale{0.6}
\ifx\JPicScale\undefined\def\JPicScale{1}\fi
\unitlength \JPicScale mm
\begin{picture}(120,70)(0,0)
\linethickness{0.3mm}
\multiput(10,50)(0.12,0.12){167}{\line(1,0){0.12}}
\linethickness{0.3mm}
\multiput(30,70)(0.12,-0.12){167}{\line(1,0){0.12}}
\linethickness{0.3mm}
\multiput(30,30)(0.12,0.12){167}{\line(1,0){0.12}}
\linethickness{0.3mm}
\multiput(10,50)(0.12,-0.12){167}{\line(1,0){0.12}}
\linethickness{.6mm}
\put(40,35){\line(0,1){10}}
\put(60,50){\makebox(0,0)[cc]{=}}

\linethickness{0.3mm}
\multiput(70,50)(0.12,0.12){167}{\line(1,0){0.12}}
\linethickness{0.3mm}
\multiput(90,70)(0.12,-0.12){167}{\line(1,0){0.12}}
\linethickness{0.3mm}
\multiput(70,50)(0.12,-0.12){167}{\line(1,0){0.12}}
\linethickness{0.3mm}
\put(90,30){\line(1,0){10}}
\linethickness{0.3mm}
\put(100,50){\line(1,0){10}}
\linethickness{.6mm}
\qbezier(15,20)(14.98,25.19)(16.19,30)
\qbezier(16.19,30)(17.39,34.81)(20,40)
\qbezier(20,40)(22.59,45.23)(25,46.44)
\qbezier(25,46.44)(27.41,47.64)(30,45)
\qbezier(30,45)(32.59,42.41)(35,39.41)
\qbezier(35,39.41)(37.41,36.4)(40,32.5)
\qbezier(40,32.5)(42.59,28.56)(45,28.56)
\qbezier(45,28.56)(47.41,28.56)(50,32.5)
\qbezier(50,32.5)(52.61,36.32)(53.81,45.34)
\qbezier(53.81,45.34)(55.02,54.37)(55,70)

\linethickness{.6mm}
\qbezier(65,20)(65,25.2)(65,28.81)
\qbezier(65,28.81)(65,32.42)(65,35)
\qbezier(65,35)(64.96,37.6)(67.97,39.41)
\qbezier(67.97,39.41)(70.98,41.21)(77.5,42.5)
\qbezier(77.5,42.5)(83.99,43.8)(89.41,44.41)
\qbezier(89.41,44.41)(94.82,45.01)(100,45)
\qbezier(100,45)(105.2,44.98)(108.81,46.19)
\qbezier(108.81,46.19)(112.42,47.39)(115,50)
\qbezier(115,50)(117.61,52.56)(118.81,57.38)
\qbezier(118.81,57.38)(120.02,62.19)(120,70)

\end{picture}

}

\def\figAonetwoe{

\def\JPicScale{0.6}
\ifx\JPicScale\undefined\def\JPicScale{1}\fi
\unitlength \JPicScale mm
\begin{picture}(110,70)(0,0)
\linethickness{0.3mm}
\multiput(10,40)(0.12,0.12){167}{\line(1,0){0.12}}
\linethickness{0.3mm}
\multiput(30,60)(0.12,-0.12){167}{\line(1,0){0.12}}
\linethickness{0.3mm}
\multiput(30,20)(0.12,0.12){167}{\line(1,0){0.12}}
\linethickness{0.3mm}
\multiput(10,40)(0.12,-0.12){167}{\line(1,0){0.12}}
\linethickness{.6mm}
\put(18,26){\line(0,1){10}}
\linethickness{.6mm}
\qbezier(35,10)(35.01,15.2)(34.41,19.41)
\qbezier(34.41,19.41)(33.8,23.62)(32.5,27.5)
\qbezier(32.5,27.5)(31.2,31.41)(30,33.22)
\qbezier(30,33.22)(28.8,35.02)(27.5,35)
\qbezier(27.5,35)(26.2,35.02)(25,33.81)
\qbezier(25,33.81)(23.8,32.61)(22.5,30)
\qbezier(22.5,30)(21.2,27.41)(20,25)
\qbezier(20,25)(18.8,22.59)(17.5,20)
\qbezier(17.5,20)(16.22,17.38)(13.81,16.78)
\qbezier(13.81,16.78)(11.41,16.18)(7.5,17.5)
\qbezier(7.5,17.5)(3.59,18.77)(1.78,21.78)
\qbezier(1.78,21.78)(-0.02,24.79)(0,30)
\qbezier(0,30)(0,35.22)(0,37.62)
\qbezier(0,37.62)(0,40.03)(0,40)
\qbezier(0,40)(0,39.97)(0,42.38)
\qbezier(0,42.38)(0,44.78)(0,50)
\qbezier(0,50)(0,55.19)(0,60)
\qbezier(0,60)(0,64.81)(0,70)
\put(60,40){\makebox(0,0)[cc]{=}}

\linethickness{0.3mm}
\multiput(70,40)(0.12,0.12){167}{\line(1,0){0.12}}
\linethickness{0.3mm}
\multiput(90,60)(0.12,-0.12){167}{\line(1,0){0.12}}
\linethickness{0.3mm}
\put(80,20){\line(1,0){10}}
\linethickness{0.3mm}
\put(70,40){\line(1,0){10}}
\linethickness{.6mm}
\put(40,25){\line(0,1){10}}
\linethickness{0.3mm}
\multiput(90,20)(0.12,0.24){42}{\line(0,1){0.24}}
\linethickness{0.3mm}
\put(100,40){\line(1,0){10}}

\linethickness{.6mm}
\qbezier(105,10)(105.02,15.19)(103.81,20)
\qbezier(103.81,20)(102.61,24.81)(100,30)
\qbezier(100,30)(97.41,35.23)(94.41,36.44)
\qbezier(94.41,36.44)(91.4,37.64)(87.5,35)
\qbezier(87.5,35)(83.6,32.39)(80.59,31.19)
\qbezier(80.59,31.19)(77.59,29.98)(75,30)
\qbezier(75,30)(72.4,29.98)(70.59,31.19)
\qbezier(70.59,31.19)(68.79,32.39)(67.5,35)
\qbezier(67.5,35)(66.2,37.55)(65.59,43.56)
\qbezier(65.59,43.56)(64.99,49.58)(65,60)

\end{picture}

}

\def\figAonetwof{

\def\JPicScale{0.6}
\ifx\JPicScale\undefined\def\JPicScale{1}\fi
\unitlength \JPicScale mm
\begin{picture}(115,75)(0,0)
\linethickness{0.3mm}
\multiput(10,50)(0.12,0.12){167}{\line(1,0){0.12}}
\linethickness{0.3mm}
\multiput(30,70)(0.12,-0.12){167}{\line(1,0){0.12}}
\linethickness{0.3mm}
\multiput(30,30)(0.12,0.12){167}{\line(1,0){0.12}}
\linethickness{0.3mm}
\multiput(10,50)(0.12,-0.12){167}{\line(1,0){0.12}}
\linethickness{.6mm}
\put(40,35){\line(0,1){10}}
\put(60,50){\makebox(0,0)[cc]{=}}

\linethickness{0.3mm}
\multiput(70,50)(0.12,0.12){167}{\line(1,0){0.12}}
\linethickness{0.3mm}
\multiput(90,70)(0.12,-0.12){167}{\line(1,0){0.12}}
\linethickness{0.3mm}
\put(90,30){\line(1,0){10}}
\linethickness{0.3mm}
\put(100,50){\line(1,0){10}}
\linethickness{.6mm}
\qbezier(15,15)(14.98,20.19)(16.19,25)
\qbezier(16.19,25)(17.39,29.81)(20,35)
\qbezier(20,35)(22.59,40.23)(25,41.44)
\qbezier(25,41.44)(27.41,42.64)(30,40)
\qbezier(30,40)(32.59,37.41)(35,34.41)
\qbezier(35,34.41)(37.41,31.4)(40,27.5)
\qbezier(40,27.5)(42.59,23.56)(45,23.56)
\qbezier(45,23.56)(47.41,23.56)(50,27.5)
\qbezier(50,27.5)(52.61,31.32)(53.81,40.34)
\qbezier(53.81,40.34)(55.02,49.37)(55,65)
\linethickness{.6mm}
\put(18,36){\line(0,1){10}}
\linethickness{0.3mm}
\put(70,50){\line(1,0){10}}
\linethickness{0.3mm}
\multiput(80,40)(0.12,-0.12){83}{\line(1,0){0.12}}

\linethickness{.6mm}
\qbezier(65,20)(65,25.2)(65,28.81)
\qbezier(65,28.81)(65,32.42)(65,35)
\qbezier(65,35)(64.96,37.6)(67.97,39.41)
\qbezier(67.97,39.41)(70.98,41.21)(77.5,42.5)
\qbezier(77.5,42.5)(83.99,43.8)(89.41,44.41)
\qbezier(89.41,44.41)(94.82,45.01)(100,45)
\qbezier(100,45)(105.2,44.98)(108.81,46.19)
\qbezier(108.81,46.19)(112.42,47.39)(115,50)
\qbezier(115,50)(117.61,52.56)(118.81,57.38)
\qbezier(118.81,57.38)(120.02,62.19)(120,70)

\end{picture}

}

\def\figfulla{

\def\JPicScale{0.6}
\ifx\JPicScale\undefined\def\JPicScale{1}\fi
\unitlength \JPicScale mm
\begin{picture}(110,80)(0,0)
\linethickness{0.3mm}
\multiput(10,50)(0.12,0.12){167}{\line(1,0){0.12}}
\linethickness{0.3mm}
\multiput(30,70)(0.12,-0.12){167}{\line(1,0){0.12}}
\linethickness{0.3mm}
\multiput(30,30)(0.12,0.12){167}{\line(1,0){0.12}}
\linethickness{0.3mm}
\multiput(10,50)(0.12,-0.12){167}{\line(1,0){0.12}}
\linethickness{0.3mm}
\linethickness{.6mm}
\put(25,25){\line(0,1){50}}
\linethickness{0.3mm}
\multiput(70,50)(0.12,0.12){167}{\line(1,0){0.12}}
\linethickness{0.3mm}
\multiput(90,70)(0.12,-0.12){167}{\line(1,0){0.12}}
\linethickness{0.3mm}
\multiput(90,30)(0.12,0.12){167}{\line(1,0){0.12}}
\linethickness{0.3mm}
\multiput(70,50)(0.12,-0.12){167}{\line(1,0){0.12}}
\linethickness{0.3mm}
\linethickness{.6mm}
\qbezier(100,80)(100.03,74.84)(97.62,67.62)
\qbezier(97.62,67.62)(95.22,60.41)(90,50)
\qbezier(90,50)(84.78,39.58)(82.38,33.56)
\qbezier(82.38,33.56)(79.97,27.55)(80,25)
\qbezier(80,25)(80,22.39)(80,21.19)
\qbezier(80,21.19)(80,19.98)(80,20)
\put(60,50){\makebox(0,0)[cc]{+}}

\put(120,50){\makebox(0,0)[cc]{+}}

\end{picture}

}

\def\figfullb{

\def\JPicScale{0.6}
\ifx\JPicScale\undefined\def\JPicScale{1}\fi
\unitlength \JPicScale mm
\begin{picture}(110,80)(0,0)
\linethickness{0.3mm}
\multiput(10,50)(0.12,0.12){167}{\line(1,0){0.12}}
\linethickness{0.3mm}
\multiput(30,70)(0.12,-0.12){167}{\line(1,0){0.12}}
\linethickness{0.3mm}
\multiput(10,50)(0.12,-0.12){167}{\line(1,0){0.12}}
\linethickness{.6mm}
\put(35,25){\line(0,1){50}}
\linethickness{0.3mm}
\multiput(70,50)(0.12,0.12){167}{\line(1,0){0.12}}
\linethickness{0.3mm}
\multiput(90,70)(0.12,-0.12){167}{\line(1,0){0.12}}
\linethickness{0.3mm}
\multiput(70,50)(0.12,-0.12){167}{\line(1,0){0.12}}
\put(60,50){\makebox(0,0)[cc]{+}}

\linethickness{.6mm}
\qbezier(75,80)(74.98,74.8)(76.78,70.59)
\qbezier(76.78,70.59)(78.59,66.38)(82.5,62.5)
\qbezier(82.5,62.5)(86.41,58.62)(88.81,54.41)
\qbezier(88.81,54.41)(91.22,50.2)(92.5,45)
\qbezier(92.5,45)(93.8,39.83)(94.41,33.81)
\qbezier(94.41,33.81)(95.01,27.8)(95,20)
\linethickness{0.3mm}
\multiput(30,30)(0.12,0.12){167}{\line(1,0){0.12}}
\linethickness{0.3mm}
\multiput(90,30)(0.12,0.12){167}{\line(1,0){0.12}}
\end{picture}

}

\def\figtreeone{

\def\JPicScale{0.6}
\ifx\JPicScale\undefined\def\JPicScale{1}\fi
\unitlength \JPicScale mm
\begin{picture}(80,60)(0,0)
\linethickness{0.3mm}
\multiput(10,60)(0.12,-0.12){167}{\line(1,0){0.12}}
\linethickness{0.3mm}
\multiput(10,20)(0.12,0.12){167}{\line(1,0){0.12}}
\linethickness{0.3mm}
\put(10,40){\line(1,0){20}}
\linethickness{0.3mm}
\put(30,40){\line(1,0){30}}
\linethickness{0.3mm}
\multiput(60,40)(0.12,0.12){167}{\line(1,0){0.12}}
\linethickness{0.3mm}
\put(60,40){\line(1,0){20}}
\linethickness{0.3mm}
\multiput(60,40)(0.12,-0.12){167}{\line(1,0){0.12}}
\put(40,35){\makebox(0,0)[cc]{$p\rightarrow$}}

\put(40,15){\makebox(0,0)[cc]{(a)}}

\end{picture}

}

\def\figtreeoneb{

\def\JPicScale{0.6}
\ifx\JPicScale\undefined\def\JPicScale{1}\fi
\unitlength \JPicScale mm
\begin{picture}(80,60)(0,0)
\linethickness{0.3mm}
\multiput(10,60)(0.12,-0.12){167}{\line(1,0){0.12}}
\linethickness{0.3mm}
\multiput(10,20)(0.12,0.12){167}{\line(1,0){0.12}}
\linethickness{0.3mm}
\put(10,40){\line(1,0){20}}
\linethickness{0.3mm}
\put(30,40){\line(1,0){30}}
\linethickness{0.3mm}
\multiput(60,40)(0.12,0.12){167}{\line(1,0){0.12}}
\linethickness{0.3mm}
\put(60,40){\line(1,0){20}}
\linethickness{0.3mm}
\multiput(60,40)(0.12,-0.12){167}{\line(1,0){0.12}}

\linethickness{.6mm}
\put(40,35){\makebox(0,0)[cc]{$p\rightarrow$}}

\put(40,15){\makebox(0,0)[cc]{(b)}}

\put(50,20){\line(0,1){40}}

\end{picture}

}

\def\figtreetwo{

\ifx\JPicScale\undefined\def\JPicScale{1}\fi
\unitlength \JPicScale mm
\begin{picture}(62.5,67.5)(0,0)
\linethickness{0.3mm}
\put(62.5,54.75){\line(0,1){0.5}}
\multiput(62.48,55.75)(0.02,-0.5){1}{\line(0,-1){0.5}}
\multiput(62.44,56.26)(0.04,-0.5){1}{\line(0,-1){0.5}}
\multiput(62.38,56.76)(0.06,-0.5){1}{\line(0,-1){0.5}}
\multiput(62.3,57.25)(0.08,-0.5){1}{\line(0,-1){0.5}}
\multiput(62.19,57.75)(0.1,-0.49){1}{\line(0,-1){0.49}}
\multiput(62.07,58.24)(0.12,-0.49){1}{\line(0,-1){0.49}}
\multiput(61.93,58.72)(0.14,-0.48){1}{\line(0,-1){0.48}}
\multiput(61.77,59.2)(0.16,-0.48){1}{\line(0,-1){0.48}}
\multiput(61.6,59.67)(0.18,-0.47){1}{\line(0,-1){0.47}}
\multiput(61.4,60.13)(0.1,-0.23){2}{\line(0,-1){0.23}}
\multiput(61.18,60.58)(0.11,-0.23){2}{\line(0,-1){0.23}}
\multiput(60.95,61.03)(0.12,-0.22){2}{\line(0,-1){0.22}}
\multiput(60.7,61.47)(0.13,-0.22){2}{\line(0,-1){0.22}}
\multiput(60.43,61.89)(0.13,-0.21){2}{\line(0,-1){0.21}}
\multiput(60.14,62.31)(0.14,-0.21){2}{\line(0,-1){0.21}}
\multiput(59.84,62.71)(0.1,-0.13){3}{\line(0,-1){0.13}}
\multiput(59.52,63.1)(0.11,-0.13){3}{\line(0,-1){0.13}}
\multiput(59.19,63.48)(0.11,-0.13){3}{\line(0,-1){0.13}}
\multiput(58.84,63.84)(0.12,-0.12){3}{\line(0,-1){0.12}}
\multiput(58.48,64.19)(0.12,-0.12){3}{\line(1,0){0.12}}
\multiput(58.1,64.52)(0.13,-0.11){3}{\line(1,0){0.13}}
\multiput(57.71,64.84)(0.13,-0.11){3}{\line(1,0){0.13}}
\multiput(57.31,65.14)(0.13,-0.1){3}{\line(1,0){0.13}}
\multiput(56.89,65.43)(0.21,-0.14){2}{\line(1,0){0.21}}
\multiput(56.47,65.7)(0.21,-0.13){2}{\line(1,0){0.21}}
\multiput(56.03,65.95)(0.22,-0.13){2}{\line(1,0){0.22}}
\multiput(55.58,66.18)(0.22,-0.12){2}{\line(1,0){0.22}}
\multiput(55.13,66.4)(0.23,-0.11){2}{\line(1,0){0.23}}
\multiput(54.67,66.6)(0.23,-0.1){2}{\line(1,0){0.23}}
\multiput(54.2,66.77)(0.47,-0.18){1}{\line(1,0){0.47}}
\multiput(53.72,66.93)(0.48,-0.16){1}{\line(1,0){0.48}}
\multiput(53.24,67.07)(0.48,-0.14){1}{\line(1,0){0.48}}
\multiput(52.75,67.19)(0.49,-0.12){1}{\line(1,0){0.49}}
\multiput(52.25,67.3)(0.49,-0.1){1}{\line(1,0){0.49}}
\multiput(51.76,67.38)(0.5,-0.08){1}{\line(1,0){0.5}}
\multiput(51.26,67.44)(0.5,-0.06){1}{\line(1,0){0.5}}
\multiput(50.75,67.48)(0.5,-0.04){1}{\line(1,0){0.5}}
\multiput(50.25,67.5)(0.5,-0.02){1}{\line(1,0){0.5}}
\put(49.75,67.5){\line(1,0){0.5}}
\multiput(49.25,67.48)(0.5,0.02){1}{\line(1,0){0.5}}
\multiput(48.74,67.44)(0.5,0.04){1}{\line(1,0){0.5}}
\multiput(48.24,67.38)(0.5,0.06){1}{\line(1,0){0.5}}
\multiput(47.75,67.3)(0.5,0.08){1}{\line(1,0){0.5}}
\multiput(47.25,67.19)(0.49,0.1){1}{\line(1,0){0.49}}
\multiput(46.76,67.07)(0.49,0.12){1}{\line(1,0){0.49}}
\multiput(46.28,66.93)(0.48,0.14){1}{\line(1,0){0.48}}
\multiput(45.8,66.77)(0.48,0.16){1}{\line(1,0){0.48}}
\multiput(45.33,66.6)(0.47,0.18){1}{\line(1,0){0.47}}
\multiput(44.87,66.4)(0.23,0.1){2}{\line(1,0){0.23}}
\multiput(44.42,66.18)(0.23,0.11){2}{\line(1,0){0.23}}
\multiput(43.97,65.95)(0.22,0.12){2}{\line(1,0){0.22}}
\multiput(43.53,65.7)(0.22,0.13){2}{\line(1,0){0.22}}
\multiput(43.11,65.43)(0.21,0.13){2}{\line(1,0){0.21}}
\multiput(42.69,65.14)(0.21,0.14){2}{\line(1,0){0.21}}
\multiput(42.29,64.84)(0.13,0.1){3}{\line(1,0){0.13}}
\multiput(41.9,64.52)(0.13,0.11){3}{\line(1,0){0.13}}
\multiput(41.52,64.19)(0.13,0.11){3}{\line(1,0){0.13}}
\multiput(41.16,63.84)(0.12,0.12){3}{\line(1,0){0.12}}
\multiput(40.81,63.48)(0.12,0.12){3}{\line(0,1){0.12}}
\multiput(40.48,63.1)(0.11,0.13){3}{\line(0,1){0.13}}
\multiput(40.16,62.71)(0.11,0.13){3}{\line(0,1){0.13}}
\multiput(39.86,62.31)(0.1,0.13){3}{\line(0,1){0.13}}
\multiput(39.57,61.89)(0.14,0.21){2}{\line(0,1){0.21}}
\multiput(39.3,61.47)(0.13,0.21){2}{\line(0,1){0.21}}
\multiput(39.05,61.03)(0.13,0.22){2}{\line(0,1){0.22}}
\multiput(38.82,60.58)(0.12,0.22){2}{\line(0,1){0.22}}
\multiput(38.6,60.13)(0.11,0.23){2}{\line(0,1){0.23}}
\multiput(38.4,59.67)(0.1,0.23){2}{\line(0,1){0.23}}
\multiput(38.23,59.2)(0.18,0.47){1}{\line(0,1){0.47}}
\multiput(38.07,58.72)(0.16,0.48){1}{\line(0,1){0.48}}
\multiput(37.93,58.24)(0.14,0.48){1}{\line(0,1){0.48}}
\multiput(37.81,57.75)(0.12,0.49){1}{\line(0,1){0.49}}
\multiput(37.7,57.25)(0.1,0.49){1}{\line(0,1){0.49}}
\multiput(37.62,56.76)(0.08,0.5){1}{\line(0,1){0.5}}
\multiput(37.56,56.26)(0.06,0.5){1}{\line(0,1){0.5}}
\multiput(37.52,55.75)(0.04,0.5){1}{\line(0,1){0.5}}
\multiput(37.5,55.25)(0.02,0.5){1}{\line(0,1){0.5}}
\put(37.5,54.75){\line(0,1){0.5}}
\multiput(37.5,54.75)(0.02,-0.5){1}{\line(0,-1){0.5}}
\multiput(37.52,54.25)(0.04,-0.5){1}{\line(0,-1){0.5}}
\multiput(37.56,53.74)(0.06,-0.5){1}{\line(0,-1){0.5}}
\multiput(37.62,53.24)(0.08,-0.5){1}{\line(0,-1){0.5}}
\multiput(37.7,52.75)(0.1,-0.49){1}{\line(0,-1){0.49}}
\multiput(37.81,52.25)(0.12,-0.49){1}{\line(0,-1){0.49}}
\multiput(37.93,51.76)(0.14,-0.48){1}{\line(0,-1){0.48}}
\multiput(38.07,51.28)(0.16,-0.48){1}{\line(0,-1){0.48}}
\multiput(38.23,50.8)(0.18,-0.47){1}{\line(0,-1){0.47}}
\multiput(38.4,50.33)(0.1,-0.23){2}{\line(0,-1){0.23}}
\multiput(38.6,49.87)(0.11,-0.23){2}{\line(0,-1){0.23}}
\multiput(38.82,49.42)(0.12,-0.22){2}{\line(0,-1){0.22}}
\multiput(39.05,48.97)(0.13,-0.22){2}{\line(0,-1){0.22}}
\multiput(39.3,48.53)(0.13,-0.21){2}{\line(0,-1){0.21}}
\multiput(39.57,48.11)(0.14,-0.21){2}{\line(0,-1){0.21}}
\multiput(39.86,47.69)(0.1,-0.13){3}{\line(0,-1){0.13}}
\multiput(40.16,47.29)(0.11,-0.13){3}{\line(0,-1){0.13}}
\multiput(40.48,46.9)(0.11,-0.13){3}{\line(0,-1){0.13}}
\multiput(40.81,46.52)(0.12,-0.12){3}{\line(0,-1){0.12}}
\multiput(41.16,46.16)(0.12,-0.12){3}{\line(1,0){0.12}}
\multiput(41.52,45.81)(0.13,-0.11){3}{\line(1,0){0.13}}
\multiput(41.9,45.48)(0.13,-0.11){3}{\line(1,0){0.13}}
\multiput(42.29,45.16)(0.13,-0.1){3}{\line(1,0){0.13}}
\multiput(42.69,44.86)(0.21,-0.14){2}{\line(1,0){0.21}}
\multiput(43.11,44.57)(0.21,-0.13){2}{\line(1,0){0.21}}
\multiput(43.53,44.3)(0.22,-0.13){2}{\line(1,0){0.22}}
\multiput(43.97,44.05)(0.22,-0.12){2}{\line(1,0){0.22}}
\multiput(44.42,43.82)(0.23,-0.11){2}{\line(1,0){0.23}}
\multiput(44.87,43.6)(0.23,-0.1){2}{\line(1,0){0.23}}
\multiput(45.33,43.4)(0.47,-0.18){1}{\line(1,0){0.47}}
\multiput(45.8,43.23)(0.48,-0.16){1}{\line(1,0){0.48}}
\multiput(46.28,43.07)(0.48,-0.14){1}{\line(1,0){0.48}}
\multiput(46.76,42.93)(0.49,-0.12){1}{\line(1,0){0.49}}
\multiput(47.25,42.81)(0.49,-0.1){1}{\line(1,0){0.49}}
\multiput(47.75,42.7)(0.5,-0.08){1}{\line(1,0){0.5}}
\multiput(48.24,42.62)(0.5,-0.06){1}{\line(1,0){0.5}}
\multiput(48.74,42.56)(0.5,-0.04){1}{\line(1,0){0.5}}
\multiput(49.25,42.52)(0.5,-0.02){1}{\line(1,0){0.5}}
\put(49.75,42.5){\line(1,0){0.5}}
\multiput(50.25,42.5)(0.5,0.02){1}{\line(1,0){0.5}}
\multiput(50.75,42.52)(0.5,0.04){1}{\line(1,0){0.5}}
\multiput(51.26,42.56)(0.5,0.06){1}{\line(1,0){0.5}}
\multiput(51.76,42.62)(0.5,0.08){1}{\line(1,0){0.5}}
\multiput(52.25,42.7)(0.49,0.1){1}{\line(1,0){0.49}}
\multiput(52.75,42.81)(0.49,0.12){1}{\line(1,0){0.49}}
\multiput(53.24,42.93)(0.48,0.14){1}{\line(1,0){0.48}}
\multiput(53.72,43.07)(0.48,0.16){1}{\line(1,0){0.48}}
\multiput(54.2,43.23)(0.47,0.18){1}{\line(1,0){0.47}}
\multiput(54.67,43.4)(0.23,0.1){2}{\line(1,0){0.23}}
\multiput(55.13,43.6)(0.23,0.11){2}{\line(1,0){0.23}}
\multiput(55.58,43.82)(0.22,0.12){2}{\line(1,0){0.22}}
\multiput(56.03,44.05)(0.22,0.13){2}{\line(1,0){0.22}}
\multiput(56.47,44.3)(0.21,0.13){2}{\line(1,0){0.21}}
\multiput(56.89,44.57)(0.21,0.14){2}{\line(1,0){0.21}}
\multiput(57.31,44.86)(0.13,0.1){3}{\line(1,0){0.13}}
\multiput(57.71,45.16)(0.13,0.11){3}{\line(1,0){0.13}}
\multiput(58.1,45.48)(0.13,0.11){3}{\line(1,0){0.13}}
\multiput(58.48,45.81)(0.12,0.12){3}{\line(1,0){0.12}}
\multiput(58.84,46.16)(0.12,0.12){3}{\line(0,1){0.12}}
\multiput(59.19,46.52)(0.11,0.13){3}{\line(0,1){0.13}}
\multiput(59.52,46.9)(0.11,0.13){3}{\line(0,1){0.13}}
\multiput(59.84,47.29)(0.1,0.13){3}{\line(0,1){0.13}}
\multiput(60.14,47.69)(0.14,0.21){2}{\line(0,1){0.21}}
\multiput(60.43,48.11)(0.13,0.21){2}{\line(0,1){0.21}}
\multiput(60.7,48.53)(0.13,0.22){2}{\line(0,1){0.22}}
\multiput(60.95,48.97)(0.12,0.22){2}{\line(0,1){0.22}}
\multiput(61.18,49.42)(0.11,0.23){2}{\line(0,1){0.23}}
\multiput(61.4,49.87)(0.1,0.23){2}{\line(0,1){0.23}}
\multiput(61.6,50.33)(0.18,0.47){1}{\line(0,1){0.47}}
\multiput(61.77,50.8)(0.16,0.48){1}{\line(0,1){0.48}}
\multiput(61.93,51.28)(0.14,0.48){1}{\line(0,1){0.48}}
\multiput(62.07,51.76)(0.12,0.49){1}{\line(0,1){0.49}}
\multiput(62.19,52.25)(0.1,0.49){1}{\line(0,1){0.49}}
\multiput(62.3,52.75)(0.08,0.5){1}{\line(0,1){0.5}}
\multiput(62.38,53.24)(0.06,0.5){1}{\line(0,1){0.5}}
\multiput(62.44,53.74)(0.04,0.5){1}{\line(0,1){0.5}}
\multiput(62.48,54.25)(0.02,0.5){1}{\line(0,1){0.5}}

\linethickness{0.3mm}
\put(62.5,29.75){\line(0,1){0.5}}
\multiput(62.48,30.75)(0.02,-0.5){1}{\line(0,-1){0.5}}
\multiput(62.44,31.26)(0.04,-0.5){1}{\line(0,-1){0.5}}
\multiput(62.38,31.76)(0.06,-0.5){1}{\line(0,-1){0.5}}
\multiput(62.3,32.25)(0.08,-0.5){1}{\line(0,-1){0.5}}
\multiput(62.19,32.75)(0.1,-0.49){1}{\line(0,-1){0.49}}
\multiput(62.07,33.24)(0.12,-0.49){1}{\line(0,-1){0.49}}
\multiput(61.93,33.72)(0.14,-0.48){1}{\line(0,-1){0.48}}
\multiput(61.77,34.2)(0.16,-0.48){1}{\line(0,-1){0.48}}
\multiput(61.6,34.67)(0.18,-0.47){1}{\line(0,-1){0.47}}
\multiput(61.4,35.13)(0.1,-0.23){2}{\line(0,-1){0.23}}
\multiput(61.18,35.58)(0.11,-0.23){2}{\line(0,-1){0.23}}
\multiput(60.95,36.03)(0.12,-0.22){2}{\line(0,-1){0.22}}
\multiput(60.7,36.47)(0.13,-0.22){2}{\line(0,-1){0.22}}
\multiput(60.43,36.89)(0.13,-0.21){2}{\line(0,-1){0.21}}
\multiput(60.14,37.31)(0.14,-0.21){2}{\line(0,-1){0.21}}
\multiput(59.84,37.71)(0.1,-0.13){3}{\line(0,-1){0.13}}
\multiput(59.52,38.1)(0.11,-0.13){3}{\line(0,-1){0.13}}
\multiput(59.19,38.48)(0.11,-0.13){3}{\line(0,-1){0.13}}
\multiput(58.84,38.84)(0.12,-0.12){3}{\line(0,-1){0.12}}
\multiput(58.48,39.19)(0.12,-0.12){3}{\line(1,0){0.12}}
\multiput(58.1,39.52)(0.13,-0.11){3}{\line(1,0){0.13}}
\multiput(57.71,39.84)(0.13,-0.11){3}{\line(1,0){0.13}}
\multiput(57.31,40.14)(0.13,-0.1){3}{\line(1,0){0.13}}
\multiput(56.89,40.43)(0.21,-0.14){2}{\line(1,0){0.21}}
\multiput(56.47,40.7)(0.21,-0.13){2}{\line(1,0){0.21}}
\multiput(56.03,40.95)(0.22,-0.13){2}{\line(1,0){0.22}}
\multiput(55.58,41.18)(0.22,-0.12){2}{\line(1,0){0.22}}
\multiput(55.13,41.4)(0.23,-0.11){2}{\line(1,0){0.23}}
\multiput(54.67,41.6)(0.23,-0.1){2}{\line(1,0){0.23}}
\multiput(54.2,41.77)(0.47,-0.18){1}{\line(1,0){0.47}}
\multiput(53.72,41.93)(0.48,-0.16){1}{\line(1,0){0.48}}
\multiput(53.24,42.07)(0.48,-0.14){1}{\line(1,0){0.48}}
\multiput(52.75,42.19)(0.49,-0.12){1}{\line(1,0){0.49}}
\multiput(52.25,42.3)(0.49,-0.1){1}{\line(1,0){0.49}}
\multiput(51.76,42.38)(0.5,-0.08){1}{\line(1,0){0.5}}
\multiput(51.26,42.44)(0.5,-0.06){1}{\line(1,0){0.5}}
\multiput(50.75,42.48)(0.5,-0.04){1}{\line(1,0){0.5}}
\multiput(50.25,42.5)(0.5,-0.02){1}{\line(1,0){0.5}}
\put(49.75,42.5){\line(1,0){0.5}}
\multiput(49.25,42.48)(0.5,0.02){1}{\line(1,0){0.5}}
\multiput(48.74,42.44)(0.5,0.04){1}{\line(1,0){0.5}}
\multiput(48.24,42.38)(0.5,0.06){1}{\line(1,0){0.5}}
\multiput(47.75,42.3)(0.5,0.08){1}{\line(1,0){0.5}}
\multiput(47.25,42.19)(0.49,0.1){1}{\line(1,0){0.49}}
\multiput(46.76,42.07)(0.49,0.12){1}{\line(1,0){0.49}}
\multiput(46.28,41.93)(0.48,0.14){1}{\line(1,0){0.48}}
\multiput(45.8,41.77)(0.48,0.16){1}{\line(1,0){0.48}}
\multiput(45.33,41.6)(0.47,0.18){1}{\line(1,0){0.47}}
\multiput(44.87,41.4)(0.23,0.1){2}{\line(1,0){0.23}}
\multiput(44.42,41.18)(0.23,0.11){2}{\line(1,0){0.23}}
\multiput(43.97,40.95)(0.22,0.12){2}{\line(1,0){0.22}}
\multiput(43.53,40.7)(0.22,0.13){2}{\line(1,0){0.22}}
\multiput(43.11,40.43)(0.21,0.13){2}{\line(1,0){0.21}}
\multiput(42.69,40.14)(0.21,0.14){2}{\line(1,0){0.21}}
\multiput(42.29,39.84)(0.13,0.1){3}{\line(1,0){0.13}}
\multiput(41.9,39.52)(0.13,0.11){3}{\line(1,0){0.13}}
\multiput(41.52,39.19)(0.13,0.11){3}{\line(1,0){0.13}}
\multiput(41.16,38.84)(0.12,0.12){3}{\line(1,0){0.12}}
\multiput(40.81,38.48)(0.12,0.12){3}{\line(0,1){0.12}}
\multiput(40.48,38.1)(0.11,0.13){3}{\line(0,1){0.13}}
\multiput(40.16,37.71)(0.11,0.13){3}{\line(0,1){0.13}}
\multiput(39.86,37.31)(0.1,0.13){3}{\line(0,1){0.13}}
\multiput(39.57,36.89)(0.14,0.21){2}{\line(0,1){0.21}}
\multiput(39.3,36.47)(0.13,0.21){2}{\line(0,1){0.21}}
\multiput(39.05,36.03)(0.13,0.22){2}{\line(0,1){0.22}}
\multiput(38.82,35.58)(0.12,0.22){2}{\line(0,1){0.22}}
\multiput(38.6,35.13)(0.11,0.23){2}{\line(0,1){0.23}}
\multiput(38.4,34.67)(0.1,0.23){2}{\line(0,1){0.23}}
\multiput(38.23,34.2)(0.18,0.47){1}{\line(0,1){0.47}}
\multiput(38.07,33.72)(0.16,0.48){1}{\line(0,1){0.48}}
\multiput(37.93,33.24)(0.14,0.48){1}{\line(0,1){0.48}}
\multiput(37.81,32.75)(0.12,0.49){1}{\line(0,1){0.49}}
\multiput(37.7,32.25)(0.1,0.49){1}{\line(0,1){0.49}}
\multiput(37.62,31.76)(0.08,0.5){1}{\line(0,1){0.5}}
\multiput(37.56,31.26)(0.06,0.5){1}{\line(0,1){0.5}}
\multiput(37.52,30.75)(0.04,0.5){1}{\line(0,1){0.5}}
\multiput(37.5,30.25)(0.02,0.5){1}{\line(0,1){0.5}}
\put(37.5,29.75){\line(0,1){0.5}}
\multiput(37.5,29.75)(0.02,-0.5){1}{\line(0,-1){0.5}}
\multiput(37.52,29.25)(0.04,-0.5){1}{\line(0,-1){0.5}}
\multiput(37.56,28.74)(0.06,-0.5){1}{\line(0,-1){0.5}}
\multiput(37.62,28.24)(0.08,-0.5){1}{\line(0,-1){0.5}}
\multiput(37.7,27.75)(0.1,-0.49){1}{\line(0,-1){0.49}}
\multiput(37.81,27.25)(0.12,-0.49){1}{\line(0,-1){0.49}}
\multiput(37.93,26.76)(0.14,-0.48){1}{\line(0,-1){0.48}}
\multiput(38.07,26.28)(0.16,-0.48){1}{\line(0,-1){0.48}}
\multiput(38.23,25.8)(0.18,-0.47){1}{\line(0,-1){0.47}}
\multiput(38.4,25.33)(0.1,-0.23){2}{\line(0,-1){0.23}}
\multiput(38.6,24.87)(0.11,-0.23){2}{\line(0,-1){0.23}}
\multiput(38.82,24.42)(0.12,-0.22){2}{\line(0,-1){0.22}}
\multiput(39.05,23.97)(0.13,-0.22){2}{\line(0,-1){0.22}}
\multiput(39.3,23.53)(0.13,-0.21){2}{\line(0,-1){0.21}}
\multiput(39.57,23.11)(0.14,-0.21){2}{\line(0,-1){0.21}}
\multiput(39.86,22.69)(0.1,-0.13){3}{\line(0,-1){0.13}}
\multiput(40.16,22.29)(0.11,-0.13){3}{\line(0,-1){0.13}}
\multiput(40.48,21.9)(0.11,-0.13){3}{\line(0,-1){0.13}}
\multiput(40.81,21.52)(0.12,-0.12){3}{\line(0,-1){0.12}}
\multiput(41.16,21.16)(0.12,-0.12){3}{\line(1,0){0.12}}
\multiput(41.52,20.81)(0.13,-0.11){3}{\line(1,0){0.13}}
\multiput(41.9,20.48)(0.13,-0.11){3}{\line(1,0){0.13}}
\multiput(42.29,20.16)(0.13,-0.1){3}{\line(1,0){0.13}}
\multiput(42.69,19.86)(0.21,-0.14){2}{\line(1,0){0.21}}
\multiput(43.11,19.57)(0.21,-0.13){2}{\line(1,0){0.21}}
\multiput(43.53,19.3)(0.22,-0.13){2}{\line(1,0){0.22}}
\multiput(43.97,19.05)(0.22,-0.12){2}{\line(1,0){0.22}}
\multiput(44.42,18.82)(0.23,-0.11){2}{\line(1,0){0.23}}
\multiput(44.87,18.6)(0.23,-0.1){2}{\line(1,0){0.23}}
\multiput(45.33,18.4)(0.47,-0.18){1}{\line(1,0){0.47}}
\multiput(45.8,18.23)(0.48,-0.16){1}{\line(1,0){0.48}}
\multiput(46.28,18.07)(0.48,-0.14){1}{\line(1,0){0.48}}
\multiput(46.76,17.93)(0.49,-0.12){1}{\line(1,0){0.49}}
\multiput(47.25,17.81)(0.49,-0.1){1}{\line(1,0){0.49}}
\multiput(47.75,17.7)(0.5,-0.08){1}{\line(1,0){0.5}}
\multiput(48.24,17.62)(0.5,-0.06){1}{\line(1,0){0.5}}
\multiput(48.74,17.56)(0.5,-0.04){1}{\line(1,0){0.5}}
\multiput(49.25,17.52)(0.5,-0.02){1}{\line(1,0){0.5}}
\put(49.75,17.5){\line(1,0){0.5}}
\multiput(50.25,17.5)(0.5,0.02){1}{\line(1,0){0.5}}
\multiput(50.75,17.52)(0.5,0.04){1}{\line(1,0){0.5}}
\multiput(51.26,17.56)(0.5,0.06){1}{\line(1,0){0.5}}
\multiput(51.76,17.62)(0.5,0.08){1}{\line(1,0){0.5}}
\multiput(52.25,17.7)(0.49,0.1){1}{\line(1,0){0.49}}
\multiput(52.75,17.81)(0.49,0.12){1}{\line(1,0){0.49}}
\multiput(53.24,17.93)(0.48,0.14){1}{\line(1,0){0.48}}
\multiput(53.72,18.07)(0.48,0.16){1}{\line(1,0){0.48}}
\multiput(54.2,18.23)(0.47,0.18){1}{\line(1,0){0.47}}
\multiput(54.67,18.4)(0.23,0.1){2}{\line(1,0){0.23}}
\multiput(55.13,18.6)(0.23,0.11){2}{\line(1,0){0.23}}
\multiput(55.58,18.82)(0.22,0.12){2}{\line(1,0){0.22}}
\multiput(56.03,19.05)(0.22,0.13){2}{\line(1,0){0.22}}
\multiput(56.47,19.3)(0.21,0.13){2}{\line(1,0){0.21}}
\multiput(56.89,19.57)(0.21,0.14){2}{\line(1,0){0.21}}
\multiput(57.31,19.86)(0.13,0.1){3}{\line(1,0){0.13}}
\multiput(57.71,20.16)(0.13,0.11){3}{\line(1,0){0.13}}
\multiput(58.1,20.48)(0.13,0.11){3}{\line(1,0){0.13}}
\multiput(58.48,20.81)(0.12,0.12){3}{\line(1,0){0.12}}
\multiput(58.84,21.16)(0.12,0.12){3}{\line(0,1){0.12}}
\multiput(59.19,21.52)(0.11,0.13){3}{\line(0,1){0.13}}
\multiput(59.52,21.9)(0.11,0.13){3}{\line(0,1){0.13}}
\multiput(59.84,22.29)(0.1,0.13){3}{\line(0,1){0.13}}
\multiput(60.14,22.69)(0.14,0.21){2}{\line(0,1){0.21}}
\multiput(60.43,23.11)(0.13,0.21){2}{\line(0,1){0.21}}
\multiput(60.7,23.53)(0.13,0.22){2}{\line(0,1){0.22}}
\multiput(60.95,23.97)(0.12,0.22){2}{\line(0,1){0.22}}
\multiput(61.18,24.42)(0.11,0.23){2}{\line(0,1){0.23}}
\multiput(61.4,24.87)(0.1,0.23){2}{\line(0,1){0.23}}
\multiput(61.6,25.33)(0.18,0.47){1}{\line(0,1){0.47}}
\multiput(61.77,25.8)(0.16,0.48){1}{\line(0,1){0.48}}
\multiput(61.93,26.28)(0.14,0.48){1}{\line(0,1){0.48}}
\multiput(62.07,26.76)(0.12,0.49){1}{\line(0,1){0.49}}
\multiput(62.19,27.25)(0.1,0.49){1}{\line(0,1){0.49}}
\multiput(62.3,27.75)(0.08,0.5){1}{\line(0,1){0.5}}
\multiput(62.38,28.24)(0.06,0.5){1}{\line(0,1){0.5}}
\multiput(62.44,28.74)(0.04,0.5){1}{\line(0,1){0.5}}
\multiput(62.48,29.25)(0.02,0.5){1}{\line(0,1){0.5}}

\put(50,55){\makebox(0,0)[cc]{$U$}}

\put(50,30){\makebox(0,0)[cc]{$D$}}

\put(50,43){\makebox(0,0)[cc]{$P$}}

\put(48,8){\makebox(0,0)[cc]{$(a)$}}

\end{picture}

}

\def\figtreecut{

\def\JPicScale{0.6}
\ifx\JPicScale\undefined\def\JPicScale{1}\fi
\unitlength \JPicScale mm
\begin{picture}(205,80)(0,0)

\linethickness{0.6mm}
\qbezier(0,75)(0,64.58)(0,58.56)
\qbezier(0,58.56)(0,52.55)(0,50)
\qbezier(0,50)(-0.02,47.41)(1.19,45)
\qbezier(1.19,45)(2.39,42.59)(5,40)
\qbezier(5,40)(7.61,37.41)(8.81,35)
\qbezier(8.81,35)(10.02,32.59)(10,30)
\qbezier(10,30)(10,27.41)(10,24.41)
\qbezier(10,24.41)(10,21.4)(10,17.5)
\qbezier(10,17.5)(10,13.6)(10,10.59)
\qbezier(10,10.59)(10,7.59)(10,5)
\linethickness{0.6mm}
\put(42,5){\line(0,1){75}}
\linethickness{0.6mm}
\qbezier(80,80)(80,72.2)(80,66.78)
\qbezier(80,66.78)(80,61.37)(80,57.5)
\qbezier(80,57.5)(80.02,53.6)(78.81,50.59)
\qbezier(78.81,50.59)(77.61,47.59)(75,45)
\qbezier(75,45)(72.39,42.41)(71.19,40)
\qbezier(71.19,40)(69.98,37.59)(70,35)
\qbezier(70,35)(70,32.41)(70,30)
\qbezier(70,30)(70,27.59)(70,25)
\qbezier(70,25)(70,22.44)(70,17.62)
\qbezier(70,17.62)(70,12.81)(70,5)
\linethickness{0.6mm}
\put(128,5){\line(0,1){75}}
\linethickness{0.6mm}
\qbezier(170,80)(170,72.2)(170,66.78)
\qbezier(170,66.78)(170,61.37)(170,57.5)
\qbezier(170,57.5)(170,53.6)(170,50.59)
\qbezier(170,50.59)(170,47.59)(170,45)
\qbezier(170,45)(170.01,42.4)(169.41,40.59)
\qbezier(169.41,40.59)(168.8,38.79)(167.5,37.5)
\qbezier(167.5,37.5)(166.2,36.21)(165,34.41)
\qbezier(165,34.41)(163.8,32.6)(162.5,30)
\qbezier(162.5,30)(161.2,27.41)(160.59,25)
\qbezier(160.59,25)(159.99,22.59)(160,20)
\qbezier(160,20)(160,17.42)(160,13.81)
\qbezier(160,13.81)(160,10.2)(160,5)
\linethickness{0.6mm}
\qbezier(195,80)(195,72.19)(195,67.38)
\qbezier(195,67.38)(195,62.56)(195,60)
\qbezier(195,60)(195,57.4)(195,55.59)
\qbezier(195,55.59)(195,53.79)(195,52.5)
\qbezier(195,52.5)(194.98,51.21)(196.19,49.41)
\qbezier(196.19,49.41)(197.39,47.6)(200,45)
\qbezier(200,45)(202.61,42.41)(203.81,40)
\qbezier(203.81,40)(205.02,37.59)(205,35)
\qbezier(205,35)(205,32.41)(205,30)
\qbezier(205,30)(205,27.59)(205,25)
\qbezier(205,25)(205,22.44)(205,17.62)
\qbezier(205,17.62)(205,12.81)(205,5)
\put(15,55){\makebox(0,0)[cc]{$U$}}

\put(15,30){\makebox(0,0)[cc]{$D$}}

\put(50,55){\makebox(0,0)[cc]{$U$}}

\put(50,30){\makebox(0,0)[cc]{$D$}}

\put(85,55){\makebox(0,0)[cc]{$U$}}

\put(85,30){\makebox(0,0)[cc]{$D$}}

\put(120,55){\makebox(0,0)[cc]{$U$}}

\put(120,30){\makebox(0,0)[cc]{$D$}}

\put(155,55){\makebox(0,0)[cc]{$U$}}

\put(155,30){\makebox(0,0)[cc]{$D$}}

\put(190,55){\makebox(0,0)[cc]{$U$}}

\put(190,30){\makebox(0,0)[cc]{$D$}}

\put(15,54.5){\circle{25}}

\put(15,31){\circle{25}}

\put(50,54.5){\circle{25}}

\put(50,31){\circle{25}}

\put(85,54.5){\circle{25}}

\put(85,31){\circle{25}}

\put(120,54.5){\circle{25}}

\put(120,31){\circle{25}}

\put(155,54.5){\circle{25}}

\put(155,31){\circle{25}}

\put(190,54.5){\circle{25}}

\put(190,31){\circle{25}}

\end{picture}

}

\def\figtreeexample{

\def\JPicScale{0.8}
\ifx\JPicScale\undefined\def\JPicScale{1}\fi
\unitlength \JPicScale mm
\begin{picture}(110,82.5)(0,0)
\linethickness{0.3mm}
\linethickness{0.3mm}
\multiput(60,60)(0.12,-0.12){83}{\line(1,0){0.12}}
\linethickness{0.3mm}
\multiput(55,75)(0.12,-0.36){42}{\line(0,-1){0.36}}
\linethickness{0.3mm}
\put(45,60){\line(1,0){15}}
\linethickness{0.3mm}
\multiput(70,50)(0.12,0.12){83}{\line(1,0){0.12}}
\linethickness{0.3mm}
\multiput(80,60)(0.36,0.12){42}{\line(1,0){0.36}}
\linethickness{0.3mm}
\put(90,70){\line(1,0){10}}
\linethickness{0.3mm}
\multiput(85,80)(0.12,-0.24){42}{\line(0,-1){0.24}}
\linethickness{0.3mm}
\multiput(60,40)(0.12,0.12){83}{\line(1,0){0.12}}
\linethickness{0.3mm}
\put(50,40){\line(1,0){10}}
\linethickness{0.3mm}
\multiput(60,40)(0.12,-0.12){83}{\line(1,0){0.12}}
\linethickness{0.3mm}
\multiput(35,45)(0.36,-0.12){42}{\line(1,0){0.36}}
\linethickness{0.3mm}
\multiput(40,30)(0.12,0.12){83}{\line(1,0){0.12}}
\linethickness{0.3mm}
\multiput(65,35)(0.24,0.12){42}{\line(1,0){0.24}}
\linethickness{0.3mm}
\put(70,30){\line(1,0){15}}
\linethickness{0.3mm}
\multiput(70,30)(0.24,-0.12){83}{\line(1,0){0.24}}
\linethickness{0.3mm}
\multiput(80,60)(0.12,0.12){83}{\line(1,0){0.12}}
\linethickness{0.3mm}
\multiput(70,50)(0.24,-0.12){83}{\line(1,0){0.24}}
\linethickness{0.3mm}
\multiput(50,30)(0.12,0.12){83}{\line(1,0){0.12}}
\linethickness{0.3mm}
\put(35,40){\line(1,0){15}}
\linethickness{0.3mm}
\put(65,35){\line(1,0){15}}
\linethickness{0.3mm}
\multiput(50,20)(0.24,0.12){83}{\line(1,0){0.24}}
\linethickness{0.3mm}
\multiput(45,70)(0.18,-0.12){83}{\line(1,0){0.18}}
\linethickness{0.3mm}
\put(80,60){\line(1,0){10}}
\linethickness{0.3mm}
\multiput(90,70)(0.24,0.12){42}{\line(1,0){0.24}}
\linethickness{0.02mm}
\put(30,15){\framebox(70,40){}}
\linethickness{0.02mm}
\put(40,47.5){\framebox(70,35){}}
\put(35,70){\makebox(0,0)[cc]{$U$}}

\put(105,20){\makebox(0,0)[cc]{$D$}}

\put(75,51){\makebox(0,0)[cc]{$P$}}

\put(65,10){\makebox(0,0)[cc]{$(b)$}}

\end{picture}

}

\def\figdistwo{

\def\JPicScale{0.5}
\ifx\JPicScale\undefined\def\JPicScale{1}\fi
\unitlength \JPicScale mm
\begin{picture}(240,90)(0,0)

\linethickness{0.6mm}
\qbezier(0,85)(-0.06,69.36)(4.75,60.94)
\qbezier(4.75,60.94)(9.56,52.52)(20,50)
\qbezier(20,50)(30.44,47.43)(35.25,43.22)
\qbezier(35.25,43.22)(40.06,39.01)(40,32.5)
\qbezier(40,32.5)(40,25.99)(40,21.78)
\qbezier(40,21.78)(40,17.57)(40,15)
\linethickness{0.6mm}
\qbezier(50,15)(47.34,33.26)(50.34,42.28)
\qbezier(50.34,42.28)(53.35,51.3)(62.5,52.5)
\qbezier(62.5,52.5)(71.62,53.76)(77.03,57.97)
\qbezier(77.03,57.97)(82.45,62.18)(85,70)
\qbezier(85,70)(87.61,77.81)(88.81,82.62)
\qbezier(88.81,82.62)(90.02,87.44)(90,90)
\linethickness{0.6mm}
\qbezier(105,90)(112.89,71.86)(111.69,53.81)
\qbezier(111.69,53.81)(110.48,35.77)(100,15)
\put(20,70){\makebox(0,0)[cc]{$B$}}

\put(20,35){\makebox(0,0)[cc]{$C$}}

\put(65,70){\makebox(0,0)[cc]{$B$}}

\put(65,35){\makebox(0,0)[cc]{$C$}}

\put(105,70){\makebox(0,0)[cc]{$B$}}

\put(100,35){\makebox(0,0)[cc]{$C$}}

\put(20,15){\makebox(0,0)[cc]{(a)}}

\put(65,15){\makebox(0,0)[cc]{(b)}}

\put(105,15){\makebox(0,0)[cc]{(c)}}

\linethickness{0.6mm}
\qbezier(135,90)(135.02,69.12)(133.22,59.5)
\qbezier(133.22,59.5)(131.41,49.88)(127.5,50)
\qbezier(127.5,50)(123.59,50.05)(121.78,46.44)
\qbezier(121.78,46.44)(119.98,42.83)(120,35)
\qbezier(120,35)(120,27.19)(120,22.38)
\qbezier(120,22.38)(120,17.56)(120,15)
\linethickness{0.6mm}
\qbezier(160,90)(159.97,69.12)(162.38,59.5)
\qbezier(162.38,59.5)(164.78,49.88)(170,50)
\qbezier(170,50)(175.22,50.09)(177.62,42.88)
\qbezier(177.62,42.88)(180.03,35.66)(180,20)
\linethickness{0.6mm}
\qbezier(180,90)(179.98,71.74)(181.19,62.72)
\qbezier(181.19,62.72)(182.39,53.7)(185,52.5)
\qbezier(185,52.5)(187.58,51.29)(191.19,43.47)
\qbezier(191.19,43.47)(194.8,35.65)(200,20)
\linethickness{0.6mm}
\qbezier(240,90)(240.02,71.77)(238.22,60.94)
\qbezier(238.22,60.94)(236.41,50.11)(232.5,45)
\qbezier(232.5,45)(228.6,39.83)(225.59,33.81)
\qbezier(225.59,33.81)(222.59,27.8)(220,20)
\put(140,70){\makebox(0,0)[cc]{$B$}}

\put(130,35){\makebox(0,0)[cc]{$C$}}

\put(135,15){\makebox(0,0)[cc]{(d)}}

\put(165,15){\makebox(0,0)[cc]{(e)}}

\put(165,70){\makebox(0,0)[cc]{$B$}}

\put(160,35){\makebox(0,0)[cc]{$C$}}

\put(195,70){\makebox(0,0)[cc]{$B$}}

\put(190,35){\makebox(0,0)[cc]{$C$}}

\put(220,70){\makebox(0,0)[cc]{$B$}}

\put(220,35){\makebox(0,0)[cc]{$C$}}

\put(195,15){\makebox(0,0)[cc]{(f)}}

\put(225,15){\makebox(0,0)[cc]{(g)}}

\linethickness{0.3mm}

\put(20,70){\circle{25}}

\put(20,35){\circle{25}}

\put(65,70){\circle{25}}

\put(65,35){\circle{25}}

\put(105,70){\circle{25}}

\put(105,35){\circle{25}}

\put(135,70){\circle{25}}

\put(135,35){\circle{25}}

\put(165,70){\circle{25}}

\put(165,35){\circle{25}}

\put(195,70){\circle{25}}

\put(195,35){\circle{25}}

\put(225,70){\circle{25}}

\put(225,35){\circle{25}}

\end{picture}

}

\def\figdisone{

\def\JPicScale{0.5}
\ifx\JPicScale\undefined\def\JPicScale{1}\fi
\unitlength \JPicScale mm
\begin{picture}(50.83,65)(0,0)

\put(40,55){\makebox(0,0)[cc]{$B$}}

\put(40,20){\makebox(0,0)[cc]{$C$}}

\put(40,55){\circle{25}}

\put(40,20){\circle{25}}

\end{picture}

}

\def\figother{

\def\JPicScale{0.6}
\ifx\JPicScale\undefined\def\JPicScale{1}\fi
\unitlength \JPicScale mm
\begin{picture}(90,60)(0,0)
\linethickness{0.3mm}
\multiput(20,60)(0.18,-0.12){167}{\line(1,0){0.18}}
\linethickness{0.3mm}
\multiput(20,20)(0.18,0.12){167}{\line(1,0){0.18}}
\linethickness{0.3mm}
\multiput(20,60)(0.42,-0.12){167}{\line(1,0){0.42}}
\linethickness{0.3mm}
\multiput(20,20)(0.42,0.12){167}{\line(1,0){0.42}}
\put(48,47){\makebox(0,0)[cc]{1}}

\put(60,25){\makebox(0,0)[cc]{4}}

\put(48,33){\makebox(0,0)[cc]{2}}

\put(60,55){\makebox(0,0)[cc]{3}}

\put(35,45){\makebox(0,0)[cc]{$\ell$}}

\put(80,30){\makebox(0,0)[cc]{$\ell-p_A-p_B$}}

\put(25,35){\makebox(0,0)[cc]{$-p_C-\ell$}}

\put(80,50){\makebox(0,0)[cc]{$p_A-\ell$}}

\end{picture}

}

\begin{document}

\baselineskip 24pt

\begin{center}
{\Large \bf  Unitarity of the Box Diagram}

\end{center}

\vskip .6cm
\medskip

\vspace*{4.0ex}

\baselineskip=18pt

\centerline{\large \rm Roji Pius$^a$ and Ashoke Sen$^b$}

\vspace*{4.0ex}

\centerline{\large \it $^a$Perimeter Institute for Theoretical Physics} 
\centerline{\large \it  Waterloo, 
ON N2L 2Y5, Canada}

\centerline{\large \it $^b$Harish-Chandra Research Institute, HBNI}
\centerline{\large \it  Chhatnag Road, Jhusi,
Allahabad 211019, India}

\vspace*{1.0ex}
\centerline{\small E-mail:  rpius@perimeterinstitute.ca, sen@hri.res.in}

\vspace*{5.0ex}

\centerline{\bf Abstract} \bigskip

The complete
proof of cutting 
rules needed for proving perturbative unitarity of
quantum field theories usually employs the largest time equation or old
fashioned perturbation theory. None of these can be generalized to string field theory that has non-local
vertices. In arXiv:1604.01783 we gave a proof of cutting rules in string field theory, which
also provides an alternative proof of cutting 
rules in ordinary quantum field theories.  In this
note we illustrate how this works for the box diagram of $\phi^4$ field theory, avoiding the contributions
from anomalous thresholds.

\vfill \eject

\tableofcontents

\sectiono{Introduction and summary} \label{s1}

Cutkosky's diagrammatic 
analysis shows that the discontinuities of a Feynman diagram across the `normal threshold'
singularities produce the result needed for unitarity of the S-matrix\cite{Cutkosky,fowler}. 
However typically a Feynman
diagram possesses many other Landau singularities {\it e.g.} anomalous thresholds 
at one loop\cite{mandelstam,cutkosky2}
and more complicated singularities at higher loop, and there are discontinuities associated
with these singularities as well. A recent discussion on these may be found
in \cite{1610.06090,1803.08899}. 
For this reason the standard approach to proving the cutting rules needed for unitarity 
makes use of indirect methods {\it e.g.} the largest time equation\cite{veltman,diagrammar} 
or old fashioned perturbation
theory\cite{sterman} based on time ordered diagrams. A recent analysis along similar line,
suitable for dealing with vertices with finite number of time derivatives, can be found in 
\cite{1701.07052}

Unfortunately these 
approaches are not suitable for proving the cutting rules for the Feynman diagrams arising
in string field theory, since the vertices are non-local, not only in space but also in time, involving
exponentials of quadratic functions of momenta. For this reason in \cite{1604.01783}
we developed a different approach
to proving the cutting rules in such theories 
based on direct analysis of Feynman diagrams.\footnote{There have been two recent
papers\cite{1803.08827,1803.08899} on unitarity of 
non-local field theories of the kind that arise in string field theory, 
but both seem to
only focus on the analysis of discontinuity across normal thresholds.} Although originally 
developed for string field theory, this approach also gives an alternative, 
diagrammatic proof of cutting
rules in ordinary quantum field theories. Another approach
to proving unitarity in local theories 
by directly dealing with momentum space Feynman diagrams
was
suggested in \cite{1612.07148}. We suspect that this
approach is closely  related to the one described in \cite{1604.01783}, 
but the precise relation  is not clear at present.

Since  \cite{1604.01783} gave an iterative proof of the cutting rules to all
orders in perturbation theory, the analysis was necessarily somewhat abstract. In this paper 
we complement
the analysis of \cite{1604.01783} by showing how the method works for 
establishing the cutting rules for the box diagram that appears in the computation
of one loop eight point amplitude in $\phi^4$ field theory. We emphasize however that
the purpose of this note is not to prove unitarity of the box diagram for which there are many derivations.
The goal is to illustrate how the iterative all order diagrammatic proof of 
the cutting rules given in
\cite{1604.01783} works for the box diagram.

The rest of the paper is organized as follows. 
In section \ref{s1.5} we discuss some general issues that arise while trying to prove
unitarity of amplitudes written as momentum space integrals.
In section \ref{s2}, which is the main body of the paper,
we show how the method developed in \cite{1604.01783} is used to prove 
cutting rules for the box diagram of
$\phi^4$ field theory. This proof assumes the validity of cutting rules for connected and disconnected
tree diagrams. For completeness, in section \ref{s3} we give a proof of cutting rules for tree
diagrams, again by making use
of the general method described in
\cite{1604.01783}.

\sectiono{The issues} \label{s1.5}

In this section we shall briefly discuss the issues that plague the proof of unitarity directly in
momentum space. The singularities
of a Feynman diagram are associated with Landau singularities where the integrand has poles due to
certain number of internal propagators going on-shell
and furthermore the integration contour over loop momenta are pinched, i.e.\ it is not possible
to move away from the poles by deforming the integration contour in the complex loop momentum
plane. At such singularities, the integral typically has a branch cut, 
leading to a discontinuity of the
amplitude across the branch cut. 
In \cite{Cutkosky} Cutkosky gave a general formula for computing the 
discontinuity across a given threshold singularity. If a certain number of propagators go on-shell at
a singularity, then the discontinuity in the amplitude 
from the corresponding branch cut is computed by
replacing the $i/(-k^2-m^2+i\eps)$ factor in each of  these propagators by
$2\pi \delta(-k^2-m^2)$.

\begin{figure}
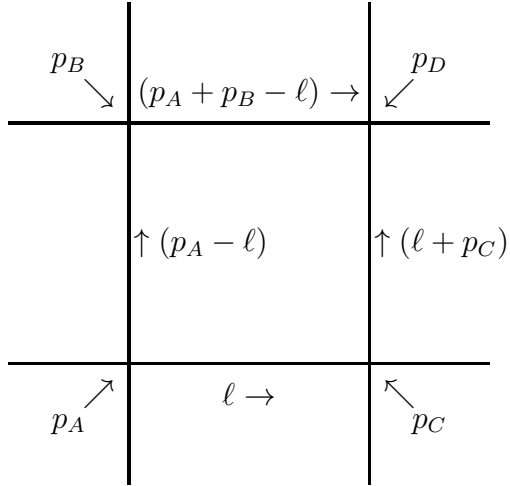


\begin{center}

\figbox

\end{center}

\caption{The box diagram of the eight point amplitude in $\phi^4$ theory. The external momenta 
$p_A,p_B,p_C,p_D$ entering at the four vertices 
are taken to be positive if ingoing.  \label{figbox}}

\end{figure}

Let us consider a singularity where the on-shell internal propagators are such that together they can be 
interpreted as 
an intermediate state in the original amplitude, {\it e.g.} if the set consists of the propagators
carrying momenta $\ell$ and $p_A+p_B-\ell$ in Fig.~\ref{figbox}.
(We are assuming that the incoming particles 
come from the left and the outgoing particles move to the right.)
Branch points associated with such singularities are known as normal thresholds.
In this case the discontinuity computed from Cutkosky's formula
can be regarded as
a product of two on-shell amplitudes, integrated over the phase space of the intermediate states.
Therefore this looks similar to a contribution to $-i\, T^\dagger T
=-i\, T^\dagger |n\rangle \langle n| T$ that is needed for unitarity of the T-matrix -- related to the
S-matrix via $S=1-i\, T$. Such
contributions are usually represented as cut diagrams, where the cut is a single line that
divides the diagram into a left half and a right half, 
with the cut propagators representing the
on-shell propagators.

While this goes a long way towards proving unitarity of the theory, there are some missing
ingradients:
\begin{enumerate}
\item A given Feynman diagram may have singularities other than normal threshold, 
{\it e.g.} if the propagators carrying momenta 
$\ell$, $\ell+p_C$ and $p_A-\ell$ in Fig.~\ref{figbox}
were on-shell at the singularity. 
Such singularities are known
as anomalous thresholds. Cutkosky's formula can still be used to compute the
discontinuity across such branch points; however
in this case the on-shell states cannot collectively be regarded
as an intermediate state in the sum $-i\, T^\dagger |n\rangle \langle n| T$.
\item In computing $-i\, T^\dagger\, T$ we need to reverse the signs of $i\eps$ in the propagators of the
amplitude to the right of the cut so that it represents a matrix element of $T^\dagger$.
This does not follow from Cutkosky's formula for discontinuity.
\end{enumerate}
While for any specific graph one can do a more detailed analysis taking into account all
these effects, the general proof of unitarity based on this approach becomes
cumbersome.
These problems were overcome in \cite{veltman,diagrammar} where a different proof of unitarity 
was given based on the `largest time equation'. An alternative proof was given in \cite{sterman} based
on the old fashioned perturbation theory. However unlike Cutkosky's original analysis, which did not
depend on the detailed structure of the vertices as long as they do not introduce additional singularities
at finite momentum, the analysis of \cite{veltman,diagrammar,sterman} requires working in 
coordinate space where at least for the time coordinate the propagator and vertices are expressed in
the position space instead of the momentum space. Unfortunately for string field theory, for which the vertices 
are exponentials of quadratic functions of momenta, there is no convenient representation of the vertices
in the coordinate space. Therefore the analysis of \cite{veltman,diagrammar,sterman} do not apply.

This difficulty was overcome in \cite{1604.01783} that expressed $T-T^\dagger$ as a
sum over cuts diagrams 
by working directly in momentum space. Furthermore the part of the contribution to the
right of the cut was shown to be hermitian conjugated, representing a contribution to 
$T^\dagger$.
This method is well-suited 
for string field theory and other non-local theories, but also for ordinary quantum field theories with
local vertices, giving an alternative proof of unitarity. In the next two sections we shall
illustrate how this works for the box diagram and tree diagrams in $\phi^4$ theory.

\sectiono{Unitarity of the box diagram} \label{s2}

Usually in $\phi^4$ theory with interaction term $-(\lambda/4!)\int d^D x \,\phi^4$ one takes the propagator
of momentum $k$ to be $-i / (k^2+m^2-i\eps)$ and the vertex to be $-i\lambda / 4!$. 
Furthermore in the computation of the T-matrix we have an overall factor of $i$.
However, as in
\cite{1604.01783}, we shall use a slightly different but equivalent
convention where for computation of the T-matrix we use the following rules:
\begin{enumerate}
\item The propagator of momentum $k$ is given by $1/(-k^2-m^2+i\eps)$.
\item The vertex is given by $\lambda/4!$.
\item For each loop integral we have a factor of $i$.
\item If the diagram has $n_c$ disconnected components then we have a factor of $i^{1-n_c}$.
\end{enumerate}
We shall drop the overall momentum conserving delta function $(2\pi)^D \delta^{(D)}\left(\sum_i p_i\right)$,
associated with each connected component,
from the expressions for the amplitudes.
The space-time dimension $D=d+1$ will be chosen such that the box diagram of Fig.~\ref{figbox} has
no ulltra-violet divergence. This requires $D\le 7$.

With this convention the contribution of the box diagram shown in Fig.~\ref{figbox} to the T-matrix
is given by,
\ben \label{eis0}
I(p_A,p_B, p_C, p_D) &=& {i\over 2} \, \lambda^4 \, 
\int {d^D \ell\over (2\pi)^D} \, \{-\ell^2-m^2+i\eps\}^{-1} \, \{ - (\ell + p_C)^2 - m^2 + i\eps\}^{-1}
\nonumber \\ &&  
\hskip -1in  
\{ - (p_A-\ell)^2 - m^2 + i\eps\}^{-1} \, \{ - (p_A+p_B-\ell)^2 - m^2 + i\eps\}^{-1} \, .
\een
$p_A$, $p_B$, $p_C$ and $p_D$ denote net external momenta entering the 
vertices.
In this convention an outgoing particle will have negative $p^0$.
Since each of $p_A$, $p_B$, $p_C$ and $p_D$ receives contribution from two
incoming or outgoing external states, they can be space-like or time-like and arbitrarily 
large in magnitude. Therefore all the singularities that can appear in the box 
diagram can be present here.
In particular by taking $p_A$ to be a large time-like momentum we can mimick the case of a
massive external particle above the threshold of production of a pair of $\phi$ particles -- this
is the situation in which the  anomalous threshold is commonly discussed.\footnote{In fact we do
not even need to assume in our analysis that the external momenta are on-shell.}

Our goal will be to compute the quantity
\be 
\DD\equiv I(p_A,p_B, p_C, p_D) - I(-p_A,-p_B, -p_C, -p_D)^*\, ,
\ee
that represents a contribution to $T-T^\dagger$.
We write down the expression for 
$I(-p_A,-p_B, -p_C, -p_D)^*$ by taking the complex conjugate of \refb{eis0} and changing the signs
of all the external momenta. Making a change of variables $\ell \to - \ell$ in the 
resulting expression,
we have
\ben \label{eis00}
I(-p_A,-p_B,-p_C, -p_D)^* &=& -{i\over 2} \, \lambda^4  \, 
\int {d^D \ell\over (2\pi)^D} \, \{-\ell^2-m^2-i\eps\}^{-1} \,
\{ - (\ell + p_C)^2 - m^2 - i\eps\}^{-1}
\nonumber \\ &&  \hskip -1.5in  
\{ - (p_A-\ell)^2 - m^2 - i\eps\}^{-1} \, \{ - (p_A+p_B-\ell)^2 - m^2 - i\eps\}^{-1} \, . 
\een
This gives
\ben \label{edd}
\DD &=& {i\over 2} \, \lambda^4 \, 
\int {d^d\ell\over (2\pi)^d}\, \int {d \ell^0 \over 2\pi}  
\Bigg[\{-\ell^2-m^2+i\eps\}^{-1}  \,
\{ - (\ell + p_C)^2 - m^2 + i\eps\}^{-1} \nonumber \\ &&
\{ - (p_A-\ell)^2 - m^2 + i\eps\}^{-1} \, \{ - (p_A+p_B-\ell)^2 - m^2 + i\eps\}^{-1} 
\nonumber \\ &&
+ \{-\ell^2-m^2-i\eps\}^{-1} \, 
\{ - (\ell + p_C)^2 - m^2 - i\eps\}^{-1} \nonumber \\ &&  
\{ - (p_A-\ell)^2 - m^2 - i\eps\}^{-1} \, \{ - (p_A+p_B-\ell)^2 - m^2 - i\eps\}^{-1} 
\bigg]\, .
\een

Unitarity of the S-matrix demands that $T-T^\dagger$ must be equal to $-i\, T^\dagger T$. This
translates to 
the cutting rules which 
tell us that $\DD$ is given by the sum over all cuts of the box diagram, with the
following rules for evaluating a cut diagram:
\begin{enumerate}
\item A cut must divide the diagram into the left half and the right half, with the convention that the
incoming particles come from the left and the outgoing particles travel to the right.
\item A cut propagator corresponds to the replacement:
\be
P(k) \equiv {1\over -k^2-m^2+i\eps} \quad \Rightarrow \quad P_c(k) \equiv - 2\, \pi\,  i\, \delta(-k^2-m^2)\, 
\theta(k^0) \, ,
\ee
where $k$ denotes the momentum flowing along the propagator from the left side of the cut to the right side.
The $-i$ factor in the expression for $P_c$ may seem unfamiliar, but in our convention this combines
with the factor of $i$ from loop integral to give the correct integration measure over the phase space.
\item The part of the amplitude to the right of the cut is replaced by its hermitian conjugate -- involving complex
conjugation and reversal of the signs of all external momenta.
\item Cut on an external line has no effect.
\item If a cut diagram has $n_L$ disconnected components on the left of the cut and $n_R$ 
disconnected
components on the right of the cut, then it should be multiplied by an additional factor of 
$(-1)^{n_R-1}$. This factor is needed to ensure that cutting rules lead to the unitarity
relation $T-T^\dagger = -i \, T^\dagger T$\cite{1604.01783}.  
\end{enumerate}

We shall first prove that for fixed $\vec\ell$ in \refb{edd}, the contribution to $\DD$ from the 
$\ell^0$ integral vanishes unless the integration contour is pinched
between two singularities.\footnote{As in \cite{1604.01783}, we shall only allow deformations of loop energy
integration contour into the complex plane, but keep the integration contours for
spatial components of the loop momenta always along the real axes.}
For this we deform the $\ell^0$ integral to $\infty$ in the lower half plane
for the first term inside the square bracket in \refb{edd}
and to $\infty$ in the upper half plane for the second term in the square bracket in \refb{edd}, 
picking up residues from
the poles that the contour passes through during the deformation. Since the poles of the first term are
complex conjugates of the poles in the second term, we pick residues from exactly the same set of
poles with $i\eps$ replaced by $-i\eps$. 
Furthermore in the first term the poles are traversed in the clockwise direction while in the
second term the poles are traversed in the anti-clockwise direction. 
As long as there are no nearby poles, we can set $\eps=0$ while evaluating these residues. 
In this case their
contributions exactly
cancel. This argument breaks down if the contours are pinched since the residues diverge as $\eps\to 0$,
and we 
have to carefully take the limit to see if there is any left-over contribution after we combine the results
of the two terms.\footnote{For string field theory the $\ell^0$ integration contour for
both terms have their ends fixed at $\pm i\infty$\cite{1604.01783}.  However a similar cancellation occurs 
for these contours as well. In this case bad behaviour in some directions at $\infty$
prevents us from deforming the contours to $\infty$, but the relevant contours can be deformed to each
other.}

The pinch singularities occur when a pair of poles in the $\ell^0$ plane approach the integration contour
from the opposite sides. Therefore in the $\vec\ell$ space they occur on a subspace of codimension 1 or higher (if more than two poles approach the same point).
We shall call this the pinched subspace.
We shall focus on the integration over a small region $\RR$ in the $\vec \ell$ space which has non-zero
intersection with the pinched subspace. We denote by $\RR'$ the image of
$\RR$ under $\vec\ell\to-\vec\ell$,  and consider
the quantity
\be 
\DD_\RR=A - A^*\, ,
\ee
where
\ben \label{edefA}
A &=& {i\over 2} \, \lambda^4 \, 
\int_\RR {d^d\ell\over (2\pi)^d}\, \int {d \ell^0 \over 2\pi} 
\{-\ell^2-m^2+i\eps\}^{-1} \,  
 \{ - (\ell + p_C)^2 - m^2 + i\eps\}^{-1}\nonumber \\  &&
\{ - (p_A-\ell)^2 - m^2 + i\eps\}^{-1} \, \{ - (p_A+p_B-\ell)^2 - m^2 + i\eps\}^{-1}\, ,
\een
and $A^*$ is obtained from $A$ by 
\begin{enumerate}
\item replacing $\RR$ by $\RR'$, 
\item reversing the signs of
all the external momenta, and
\item complex conjugation.
\end{enumerate}
We shall prove that $\DD_\RR$ is given by the sum over cuts of the contributions to $A$.
The full cutting rule is then obtained by adding the contributions from each 
small region $\RR$ of this type.

In this section we shall prove the cutting rule for $\DD_\RR$ 
assuming that it holds for all 
tree diagrams -- including disconnected ones.
This analysis will follow closely the one given in section 5.2.3 of \cite{1604.01783} for one vertex irreducible
diagrams.
In the next section we shall describe the proof of cutting rules
for connected and disconnected tree diagrams.

\begin{figure}
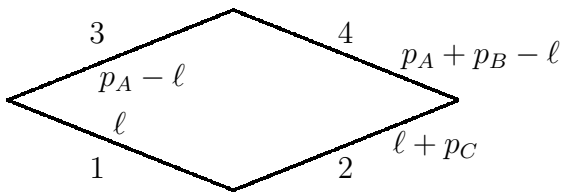


\begin{center}

\figboxtwo

\end{center}

\vskip -.8in

\caption{The reduced diagram of Fig.~\ref{figbox} associated with $\RR$ when at the pinch all 
propagators are nearly on-shell and the energies carried by the propagators lie in the range
\refb{eineq}.
The internal lines are drawn in a way so that at the pinch energy
flows from the left to the right along each of the propagators.  \label{figboxtwo}}

\end{figure}

Now for $\vec\ell\in\RR$ a certain number of propagators become nearly on-shell 
when $\ell^0$ takes the value where its integration
contour is nearly pinched. Since for small $\RR$, $\ell^0$ lies within a small range at the pinch,
we can assign definite signs 
to the energies carried by each internal propagator at the pinch. We shall now 
associate with the region $\RR$ a reduced diagram that is
obtained from the original diagram by shrinking to points all propagators that are not nearly
on-shell near the pinch. Furthermore we shall draw the nearly on-shell
propagators such that energy flows from left to right near the pinch. 
For definiteness, and to consider a situation of maximal complexity, 
we 
shall consider a region $\RR$ for which all four internal propagators are nearly on-shell at the pinch,
and\footnote{A similar analysis can be carried out for all other reduced diagrams.}
\be\label{eineq}
\ell^0>0, \quad \ell^0+p_C^0>0, \quad p_A^0-\ell^0>0, \quad p_A^0+p_B^0-\ell^0>0\, ,
\ee
at the pinch. The corresponding reduced diagram is shown in Fig.~\ref{figboxtwo}. We have
dropped the external legs from this diagram to avoid cluttering. 
We also 
number the propagators carrying momenta $\ell$, $\ell+p_C$, $p_A-\ell$ and $p_A+p_B-\ell$ by
$1$, $2$, $3$ and $4$ respectively. 
It is easy to see that the corresponding integral for $A^*$, after making a change of integration variable
$\ell^0\to -\ell^0$, will be pinched at the same value of $\ell^0$.

Let us denote by $P(k)$ the propagator with momentum $k$:
\be\label{epropfactor}
P(k)\equiv {1\over -k^2-m^2+i\eps} = {1\over k^0 -\sqrt{\vec k^2+m^2} +i\ve} \, 
{1\over k^0 +\sqrt{\vec k^2+m^2} -i\ve} 
\, ,
\ee
where $\ve$ is positive for positive $\eps$.
Therefore we can express \refb{edefA} as
\ben \label{eirp}
A&=& {i\over 2} \, \lambda^4 \,  
\int_\RR {d^d\ell\over (2\pi)^d}\, \int {d \ell^0 \over 2\pi} \, P(\ell) \, P(\ell+p_C) \, 
P(p_A-\ell)\, P(p_A+p_B-\ell)\, .
\een
Since the pinch is assumed to occur at the positive values of $\ell^0$, $\ell^0+p_C^0$,
$p_A^0-\ell^0$ and $p_A^0+p_B^0-\ell^0$, the relevant poles of the propagators
that take part in pinching the contour, are at
\ben\label{epolepos}
&& \hskip -1in \ell^0 = \sqrt{\vec\ell^2+m^2}-i\ve, \quad \ell^0=
-p_C^0 + \sqrt{(\vec\ell+\vec p_C)^2+m^2}-i\ve, 
\nonumber \\ && \hskip -1in
\ell^0=p_A^0 - \sqrt{(\vec p_A-\vec\ell)^2+m^2}+i\ve, \quad
\ell^0=p_A^0+p_B^0 - \sqrt{(\vec p_A+\vec p_B-\vec\ell)^2+m^2}+i\ve\, .
\een
Note that at the pinch the poles from the propagators 1 and 2 are in the lower half $\ell^0$
plane while the poles
from the other propagators are in the upper half $\ell^0$ 
plane. Therefore while deforming the $\ell^0$ contour to the lower half plane, we shall pick up
residues from the poles of the propagators 1 and 2 at the pinch.
For this reason the set $\{1,2\}$ will play a
special role in our analysis.

\begin{figure}
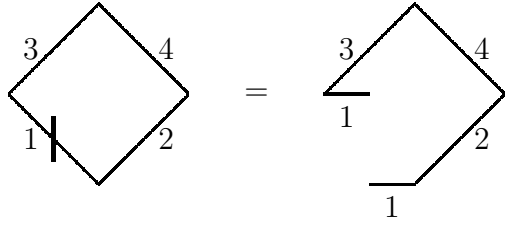


\begin{center}

\figAone

\end{center}

\vskip -.4in

\caption{Representation of $A^{(1)}$. On the left hand side the vertical line through propagator
1 represents that it is a cut propagator. On the right hand side this is made explicit by replacing the
cut propagator by a pair of external lines -- one outgoing and one incoming.  \label{figAone}}

\end{figure}

\begin{figure}
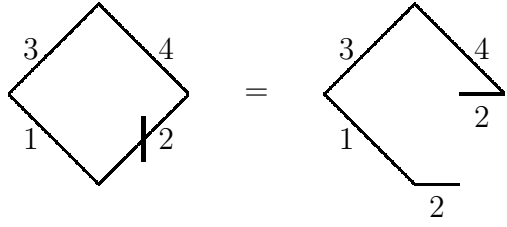


\begin{center}

\figAtwo

\end{center}

\vskip -.6in

\caption{Representation of $A^{(2)}$.   \label{figAtwo}}

\end{figure}

\begin{figure}
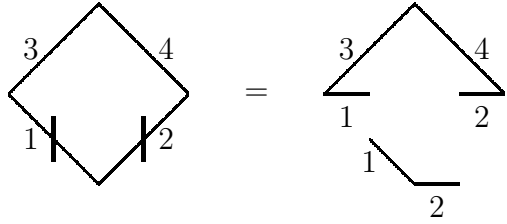


\begin{center}

\figAthree

\end{center}

\vskip -.6in

\caption{Representation of $A^{(12)}$.  \label{figAthree}}

\end{figure}

We now define
\be \label{ePdiff}
P'(k) \equiv  {1\over k^0 -\sqrt{\vec k^2+m^2} -i\ve} \, {1\over k^0 +\sqrt{\vec k^2+m^2} -i\ve} 
= P(k) - P_c(k)
\, ,
\ee
where
\be 
P_c(k) \equiv - 2\, \pi\, i\, \delta(-k^2-m^2) \, \theta(k^0)\, ,
\ee
is the cut propagator. It follows from \refb{ePdiff} that 
\be \label{edecompose}
P(\ell) P(\ell+p_C) = 
P'(\ell) P'(\ell+p_C) + P_c(\ell) P(\ell+p_C) + P(\ell) P_c(\ell+p_C) - P_c(\ell) P_c(\ell+p_C)\, .
\ee
Using this in \refb{eirp} we can express $A$ as
\be  \label{ex1}
A = \wh A + A^{(1)} + A^{(2)} - A^{(12)}\, ,
\ee
where
\ben \label{ewhA}
\wh A&=& {i\over 2} \, \lambda^4 \, 
\int_\RR {d^d\ell\over (2\pi)^d}\, \int {d \ell^0 \over 2\pi} \, P'(\ell) \, P'(\ell+p_C) \, 
P(p_A-\ell)\, P(p_A+p_B-\ell)\, ,
\een
\ben \label{eA1}
A^{(1)}&=& {i\over 2} \, \lambda^4 \, 
\int_\RR {d^d\ell\over (2\pi)^d}\, \int {d \ell^0 \over 2\pi} \, P_c(\ell) \, P(\ell+p_C) \, 
P(p_A-\ell)\, P(p_A+p_B-\ell)\, ,
\een
\ben \label{eA2}
A^{(2)}&=& {i\over 2} \, \lambda^4 \, 
\int_\RR {d^d\ell\over (2\pi)^d}\, \int {d \ell^0 \over 2\pi} \, P(\ell) \, P_c(\ell+p_C) \, 
P(p_A-\ell)\, P(p_A+p_B-\ell)\, ,
\een
\ben \label{eA12}
A^{(12)}&=& {i\over 2} \, \lambda^4 \, 
\int_\RR {d^d\ell\over (2\pi)^d}\, \int {d \ell^0 \over 2\pi} \, P_c(\ell) \, P_c(\ell+p_C) \, 
P(p_A-\ell)\, P(p_A+p_B-\ell)\, .
\een
In writing \refb{eA1}-\refb{eA12} we have used the notation of \cite{1604.01783} in 
which $A^{(i_1\cdots i_n)}$ is
obtained from $A$ by replacing the $i_1,\cdots i_n$-th propagators by cut propagators. The 
diagrammatic representations of $A^{(1)}$, $A^{(2)}$ and $A^{(12)}$ have been shown in 
Fig.~\ref{figAone}-\ref{figAthree}
with the thick vertical line across a propagator representing a cut propagator $P_c$.
We emphasize that these are not yet cut diagrams as the cut does not divide the graph into 
a left half and a right half, 
and we neither reverse the momenta not complex conjugate any part of the graph. 
Instead these should be regarded as tree amplitudes since a cut propagator
can be regarded as a pair of incoming and outgoing lines with identical momentum. By a similar 
manipulation we can express $A^*$ as
\be \label{ex2}
A^* =\wh A^* + A^{(1)*} + A^{(2)*} - A^{(12)*}\, .
\ee

\begin{figure}
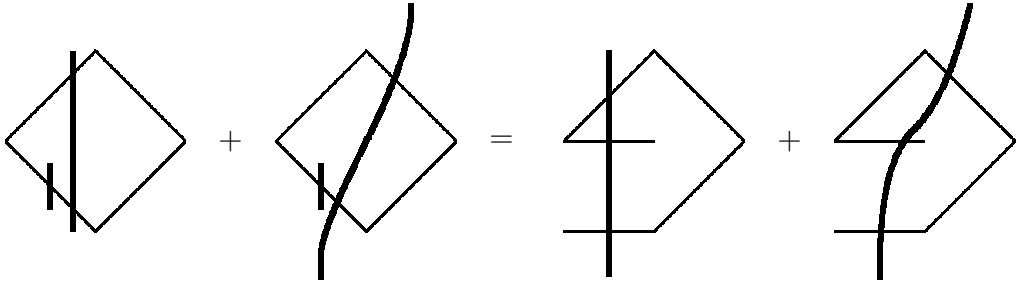


\begin{center}

\hbox{\figAonea \quad \figAoneb}

\end{center}

\vskip -.8in

\caption{Contributions to $A^{(1)}_1$. The figure on the left hand side expresses it as a sum of
cut diagrams of the original graph, with the propagator 1 replaced by the cut propagator. The
right hand side shows this as a sum of cuts of a tree diagram in which the propagator 1
is replaced by a pair of incoming and outgoing lines.
We have not shown the
external states of the original amplitude $A$ in
any of the diagrams. \label{figAoneab}}

\end{figure}

\begin{figure}
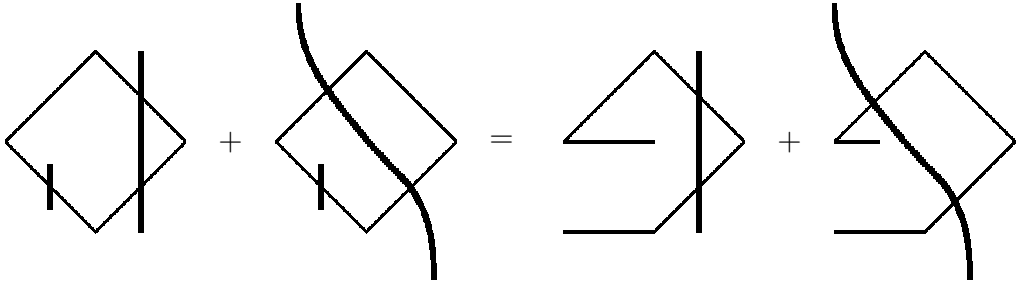


\begin{center}

\hbox{\figAonec \quad \figAoned}

\end{center}

\vskip -.8in

\caption{Contributions to $A^{(1)}_2$.  Unless cancelled, this would represent contributions
from anomalous threshold. \label{figAonecd}}

\end{figure}

\begin{figure}
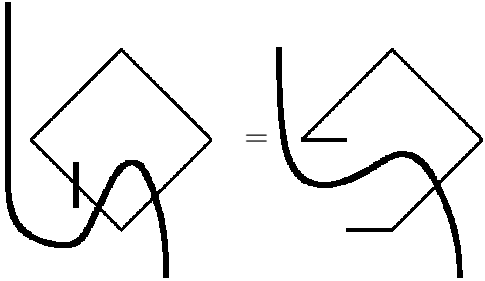


\begin{center}

\figAonee

\end{center}

\vskip -.4in

\caption{Contribution to $A^{(1)}_{\underline{1} 2}$.  In the left hand side the cut seems to 
cross the propagator 1 of the original diagram in the reverse direction so that it appears as if
the energy of the propagator 1 flows from the right of the cut to the left of the cut.
However since the propagator 1 is already on-shell, the correct representation of the diagram
is on the right hand side where it is represented as the cut of a tree diagram. In this representation
there is nothing unusual. \label{figAonee}}

\end{figure}

\begin{figure}
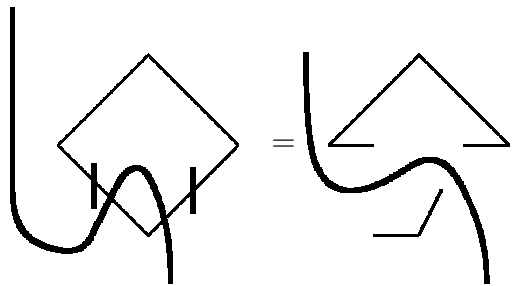


\begin{center}

\figAonetwoe 

\end{center}

\vskip -.4in

\caption{Contributions to $A^{(12)}_{\underline{1}2}$.  \label{figAonetwoe}}

\end{figure}

We now note that in the expression \refb{ewhA}
for $\wh A$ the relevant poles in the $\ell^0$ plane, responsible for the pinch in the original
amplitude $A$, are all in the
upper half plane, since in the locations of the poles of the integrand, 
the signs of the $i\ve$ in the first line of \refb{epolepos} are reversed. Therefore
the $\ell^0$ contour is not pinched, and by our previous argument,
\be \label{ex3}
\wh A-\wh A^*=0\, .
\ee
On the other hand, since $A^{(1)}$, $A^{(2)}$ and $A^{(12)}$ are tree amplitudes, and since we
are assuming that cutting rules hold for tree amplitudes, 
$A^{(1)}-A^{(1)*}$, $A^{(2)}-A^{(2)*}$ and $A^{(12)}-A^{(12)*}$ are given by sum over cuts of the
tree diagrams. This gives, in the notation of \cite{1604.01783}:
\ben \label{ex4}
&& A^{(1)}-  A^{(1)*}= A^{(1)}_1 + A^{(1)}_2 + A^{(1)}_{\underline{1} 2}\, ,
\quad A^{(2)}- A^{(2)*}=A^{(2)}_1 + A^{(2)}_2 + A^{(2)}_{\underline{2} 1}, \nonumber \\
&& A^{(12)}-
A^{(12)*}= A^{(12)}_1 +
A^{(12)}_2 + A^{(12)}_{\underline{1} 2} + A^{(12)}_{\underline{2} 1}\, .
\een
Here $A^{(i_1\cdots i_n)}_{j_1\cdots j_m}$ for $1\le i_k\le 2$, $1\le j_k\le 2$ represents sum over all cuts
of $A^{(i_1\cdots i_n)}$ satisfying the following properties:
\begin{enumerate}
\item The cut can be viewed as a cut of the original graph contributing to the amplitude $A$.
\item  The cut crosses the 
$j_1,\cdots j_m$'th propagators in the set $\{1,2\}$ and possibly other propagators outside the set $\{1,2\}$.
\end{enumerate} 
On the other hand $A^{(i_1\cdots i_n)}_{\underline{i_1} j}$ 
describes sum over cuts of $A^{(i_1\cdots i_n)}$ which pass through the $j$-th and $i_1$-th
propagators in the set $\{1,2\}$ and possibly other propagators outside the set $\{1,2\}$, 
but which are not regular cuts of the original 
amplitude $A$ since,
viewed in the context of the original graph, the $i_1$-th propagator carries
energy across the cut in the wrong direction.
Explicit diagrammatic representation of all the terms on the right hand side of \refb{ex4}
has been given in Figs.~\ref{figAoneab}-\ref{figAonetwof}. In particular 
Figs.~\ref{figAonee}, \ref{figAtwoe}, \ref{figAonetwoe} and \ref{figAonetwof} describe contributions
to $A^{(1)}_{\underline{1}2}$,  $A^{(2)}_{\underline{2}1}$, $A^{(12)}_{\underline{1}2}$ and
$A^{(12)}_{\underline{2}1}$ respectively.
As is clear from the
right hand sides of
these figures,
these are perfectly good cuts of the tree amplitude $A^{(i_1\cdots i_n)}$, even though the left hand
sides of these figures show that
they are not valid cuts of $A$.

\begin{figure}
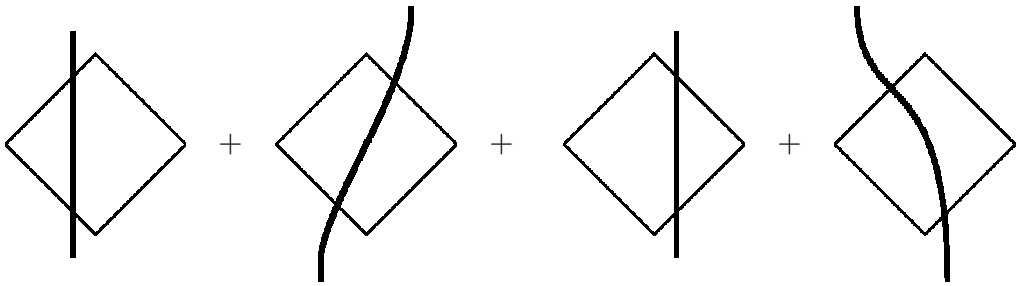


\begin{center}

\hbox{\figfulla \quad \figfullb}

\end{center}

\vskip -.8in

\caption{Complete contribution to $A-A^*$.  \label{figfin}}

\end{figure}

We shall now write down a few identities following from the simple observation that 
a propagator cut twice has the same expression as the 
propagator cut once, since the cut passing through an external line has no
effect. 
Therefore we have
\be\label{ex5}
A^{(1)}_2 = A^{(12)}_2, \quad  A^{(2)}_1 = A^{(12)}_1, \quad A^{(1)}_{\underline{1}2}
= A^{(12)}_{\underline{1}2}, \quad A^{(2)}_{\underline{2}1}
= A^{(12)}_{\underline{2}1}\, .
\ee
These identities can be verified by explicitly examining the equalities of Figs.~\ref{figAonecd} and
\ref{figAonetwocd}, \ref{figAtwoab} and \ref{figAonetwoab}, \ref{figAonee} and \ref{figAonetwoe},
and Figs.~\ref{figAtwoe} and \ref{figAonetwof}.
Using \refb{ex1}, \refb{ex2}, \refb{ex3}, \refb{ex4} and \refb{ex5} we now get
\ben
A-A^*&=& A^{(1)}_1 + A^{(1)}_2 + A^{(1)}_{\underline{1} 2} +
A^{(2)}_1 + A^{(2)}_2 + A^{(2)}_{\underline{2} 1}-A^{(12)}_1 -
A^{(12)}_2 - A^{(12)}_{\underline{1} 2} - A^{(12)}_{\underline{2} 1}\nonumber\\ &=&
A^{(1)}_1+A^{(2)}_2\, .
\een
The diagrammatic representation of the right hand side, 
given by the sum of the left hand sides of Fig.~\ref{figAoneab}
and Fig.~\ref{figAtwocd}, is shown in Fig.~\ref{figfin}.  
We now see that this is precisely given by the sum of all possible cuts of the reduced diagram 
shown in Fig.~\ref{figboxtwo}. In particular possible
contributions from anomalous thresholds, represented by 
Fig.~\ref{figAonecd} and  Fig.~\ref{figAtwoab}, cancel with the contributions from Figs.~\ref{figAonetwocd}
and \ref{figAonetwoab} and
do not appear in the final expression. These cancellations are special cases of the general results
described in eqs.(5.26)-(5.33) of \cite{1604.01783}.

\begin{figure}
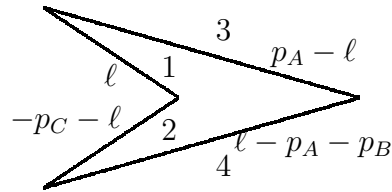


\begin{center}

\figother

\end{center}

\vskip -.4in

\caption{A reduced diagram in which non-adjacent propagators 1 and 4 
have poles on the same
side at the pinch. At the pinch the energy flows from 
left to right in each of the propagators. In the figure the momenta are labelled so that they flow from left to
right. \label{figother}}

\end{figure}

We end this section with a few remarks:
\begin{enumerate}
\item Our analysis automatically includes other reduced diagrams that are obtained by contracting
one or more propagators in Fig.~\ref{figboxtwo}. 
We simply have to set to zero all terms where the corresponding 
propagator is replaced by a cut propagator. For example if we take the triangle diagram obtained
by contracting the propagator 3 in Fig.~\ref{figboxtwo}, the final answer for $A-A^*$
will include sum over only the second and third diagrams in Fig.~\ref{figfin}.
\item This does not cover all the cases however.
An example is shown in Fig.~\ref{figother} where at the pinch 
poles on the same side come from
non-adjacent propagators. We can analyze this by repeating the analysis, with the role of the
momenta $\ell$ and $\ell+p_C$ in \refb{edecompose} now played by $\ell$ and
$\ell-p_A-p_B$. The rest of the analysis proceeds as before, with the role of the set $\{1,2\}$ played by
the set $\{1,4\}$. The final result for $A-A^*$, according to the general result of
\cite{1604.01783},  will be given by $A^{(1)}_1+A^{(4)}_4+A^{(14)}_{14}$, 
which is
simply the sum over all cuts of Fig.~\ref{figother}. (The corresponding contribution $A^{(12)}_{12}$ was
not present in the previous example since Fig.~\ref{figboxtwo} has no cut that passes 
through both propagators 1 and 2.)
\item In our analysis we have assumed that for a given $\RR$, the $\ell^0$ contour has a single 
pinch point to which two or more poles approach. We can also have more than one pinch on the $\ell^0$
contour, with two or more poles approaching each pinch point. 
Since after we factorize each propagator as
in \refb{epropfactor}, each denominator factor is linear in $\ell^0$ and has a single zero, different denominator
factors must be responsible for different pinches. Therefore 
we can divide the denominators into different
sets, with the first set $S_1$ responsible for the first pinch, 
the second set $S_2$ responsible for the
second pinch and so on. 
Different pinch points will
have different reduced diagrams associated with them, since the list of singular propagators
and the direction of energy flow through these propagators will depend on the
pinch. 

We can now carry out the analysis by first treating the product of denominators
in the set $S_1$ as in \refb{edecompose}. The main difference will be that now the $\ell^0$ contour
in $\wh A$ is still pinched due to the other set of denominators belonging to $S_2,S_3,\cdots$ etc.
The other terms contain delta functions that force $\ell^0$ to be at the first pinch and therefore the
denominators in the other sets remain finite. 
These terms can be analyzed as before.
We can now analyze $\wh A$ by decomposing the second
set of denominators, belonging to the set $S_2$, 
as in \refb{epropfactor} and repeat the analysis. We continue this
till we reach a stage where we have
a sum of terms where in one term
the contour is not pinched (the analog of $\wh A$) and in the other
terms the delta-function
fixes $\ell^0$ at a pinch.  The final result will then be given by the sum of cuts of all the reduced diagrams corresponding to all the pinches.
\end{enumerate}

\begin{figure}

\begin{center}

\hbox{\figAtwoa \quad \figAtwob}

\end{center}

\vskip -.8in

\caption{Contributions to $A^{(2)}_1$.  \label{figAtwoab}}

\end{figure}

\begin{figure}

\begin{center}

\hbox{\figAtwoc \quad \figAtwod}

\end{center}

\vskip -.8in

\caption{Contributions to $A^{(2)}_2$.  \label{figAtwocd}}

\end{figure}

\begin{figure}

\begin{center}

\figAtwoe

\end{center}

\vskip -.4in

\caption{Contributions to $A^{(2)}_{\underline{2}1}$.  \label{figAtwoe}}

\end{figure}

\begin{figure}

\begin{center}

\hbox{\figAonetwoa \quad \figAonetwob}

\end{center}

\vskip -.8in

\caption{Contributions to $A^{(12)}_1$.  \label{figAonetwoab}}

\end{figure}

\begin{figure}

\begin{center}

\hbox{\figAonetwoc \quad \figAonetwod}

\end{center}

\vskip -.8in

\caption{Contributions to $A^{(12)}_2$.  \label{figAonetwocd}}

\end{figure}

\begin{figure}

\begin{center}

\figAonetwof

\end{center}

\vskip -.4in

\caption{Contributions to $A^{(12)}_{\underline{2}1}$.  \label{figAonetwof}}

\end{figure}

\eject

\sectiono{Unitarity of tree diagrams} \label{s3}

Since the proof of cutting rules for the box diagram assumed the validity of cutting rules for
connected and disconnected tree diagrams, we shall prove the cutting rules for
tree diagrams in this section. The analysis
is a straightforward application of sections 5.2.2 and 5.3 of \cite{1604.01783}.

\begin{figure}
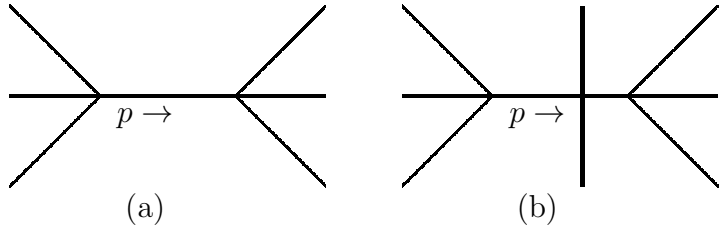


\begin{center}

\hbox{\hskip 1in \figtreeone\figtreeoneb}

\end{center}

\vskip -.4in

\caption{(a) A tree diagram in $\phi^4$ theory with six external lines and one internal propagator.
In drawing this we have made an exception to our conventions and have drawn the external lines.
The internal line has energy $p^0$ flowing from the left to the right. 
(b) Cut diagram of (a). \label{figtreeone}}

\end{figure}

We begin with the simple diagram shown in Fig.~\ref{figtreeone}(a) with
$p^0>0$. Its expression is given by
\be 
A = \lambda^2 \, (-p^2 - m^2 +i\eps)^{-1}\, .
\ee
Therefore 
\be 
A - A^* =  \lambda^2 \, (-p^2 - m^2 +i\eps)^{-1} - 
 \lambda^2 \, (-p^2 - m^2 -i\eps)^{-1}= \lambda^2  \, (-2\pi \, i)\, \delta (-p^2 - m^2) \, .
\ee
On the other hand the cut diagram shown in Fig.~\ref{figtreeone}(b) has the same expression.
(We can drop the $\theta(p^0)$ term from the cut propagator since $p^0$ has been chosen to be 
positive.)
This proves the cutting rule for Fig.\ref{figtreeone}(a).

\begin{figure}
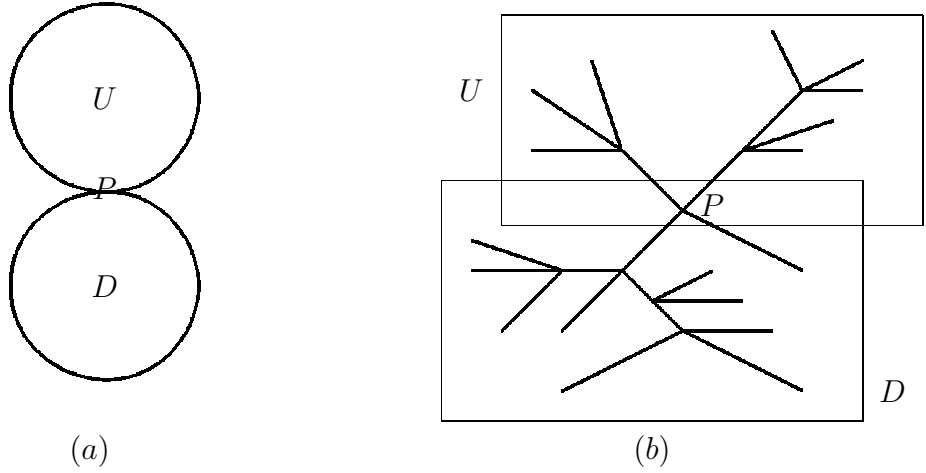


\begin{center}

\hbox{\figtreetwo \quad  \figtreeexample}

\end{center}

\vskip -.4in

\caption{(a) Schematic representation of a generic connected 
tree diagram in which the blobs $U$ and $D$ themselves are connected tree diagrams. 
(b) Example of such a generic connected tree diagram.
\label{figtreetwo}}

\end{figure}

Next we shall prove the cutting rules for any connected 
tree amplitude assuming that it holds for all connected tree amplitudes with at
least one less vertex. For this we follow closely the analysis of section 5.2.2 of
\cite{1604.01783} of `one vertex reducible' diagrams. 
Let $P$ be any vertex of the amplitude to which at least two
internal lines are connected. Then the general form of the diagram can be represented as in
Fig.~\ref{figtreetwo}(a), with each of the blobs $U$ and $D$ describing some connected tree
diagram. If $A_U$ and $A_D$ are the amplitudes associated with the tree diagrams
$U$ and $D$ respectively, then the full amplitude $A$ is given by\footnote{Note that
some of the external
lines of $U$ at $P$ are internal lines of $D$ and some of
the external lines of $D$ at $P$ are internal
lines of $U$. Therefore they are generically off-shell. This does not have any effect on our analysis 
since the validity of cutting rules does not require the external lines to be on-shell.}
\be \label{eyone}
A =\lambda^{-1} \, A_U\, A_D\, ,
\ee
where the $\lambda^{-1}$ factor accounts for the fact that both $A_U$ and $A_D$ includes a factor
of $\lambda$ from the vertex $P$ whereas in $A$ we have only one factor of $\lambda$ from the vertex.
Therefore
\be \label{eytwo}
A-A^* = \lambda^{-1} \, \{A_U\, A_D - A_U^*\, A_D^*\}\, .
\ee
Now since $A_U$ and $A_D$ are themselves connected
tree amplitudes with less number of vertices than 
$A$, $A_U-A_U^*$ and $A_D-A_D^*$ are given by sum over cut diagrams of $U$ and $D$. We divide
each of these cut diagrams into two classes: $\Delta_{UR}$ and $\Delta_{UL}$ will denote the sum over
cut diagrams
of $U$ for which the cut passes on the left and right of $P$ respectively, and similarly 
$\Delta_{DR}$ and $\Delta_{DL}$ will denote the sum over
cut diagrams
of $D$ for which the cut passes on the left and right of $P$ respectively. Therefore we have
\be\label{eythree}
A_U-A_U^* = \Delta_{UL}+\Delta_{UR}\, , \quad A_D-A_D^* = \Delta_{DL}+\Delta_{DR}\, .
\ee
Using \refb{eythree} and some trivial rearrangement of terms we can express \refb{eytwo} as
\be\label{eysix}
A-A^* =  \lambda^{-1}\left\{A_U^*\, \Delta_{DL} + 
\Delta_{UL}\Delta_{DL} + \Delta_{UL} A_D^* -\Delta_{UR}\Delta_{DR}
+ A_U \Delta_{DR} +\Delta_{UR} A_D\right\}\, .
\ee
This can be verified {\it e.g.} by expressing
both sides in terms of $A_U^*$, $A_D^*$, $\Delta_{UR}$, $\Delta_{UL}$,
$\Delta_{DR}$ and $\Delta_{DL}$.
The diagrammatic representations of the six terms in \refb{eysix}
have been shown in Fig.~\ref{figtreecut}. Special
attention should be paid to the minus sign of the fourth term on the right hand side of 
\refb{eysix}. This is compatible with the fourth term in Fig.~\ref{figtreecut} due to the
$(-1)^{n_R-1}$ factor that multiplies each cut diagram, $n_R$ being the number of disconnected
components on the right of the cut. 
If we denote by $n_{UR}$ and $n_{DR}$ the number of disconnected components on the right of the
cut in $U$ and in $D$, then the product of the cut diagrams of $U$ and $D$ carries a factor of
$(-1)^{n_{UR}-1+n_{DR}-1}$. On the other hand the fourth cut diagram of Fig.~\ref{figtreecut}
carries a factor of $(-1)^{n_{UR}+n_{DR}-1}$. The two differ by a sign showing that the
fourth diagram of Fig.~\ref{figtreecut} is indeed given by $-\lambda^{-1}\Delta_{UR} \Delta_{DR}$.

\begin{figure}

\begin{center}

\figtreecut

\end{center}

\vskip -.4in

\caption{Diagrammatic representation of the six terms in \refb{eysix}. These can also
be interpreted as cuts of Fig.~\ref{figtreetwo}(a). 
\label{figtreecut}}

\end{figure}

We now note that the six cut diagrams of Fig.~\ref{figtreecut} exhaust all possible cuts of the
diagram \ref{figtreetwo}(a). This shows that $A-A^*$ is indeed given by the sum over all cut
diagrams of
$A$ in accordance with the cutting rules.

\begin{figure}
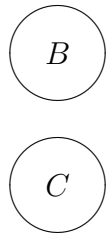


\begin{center}


\figdisone

\end{center}

\vskip -.4in

\caption{Schematic representation of a  tree amplitude with at least two disconnected components. The
blobs $B$ and $C$ represent tree amplitudes which themselves may have additional disconnected
components.
\label{figdisone}}

\end{figure}

\begin{figure}

\begin{center}


\figdistwo

\end{center}

\vskip -.4in

\caption{Diagrammatic representation of the seven terms in \refb{eseven}. These can also
be interpreted as cuts of Fig.~\ref{figdisone}. 
\label{figdistwo}}

\end{figure}

Finally we turn to the proof of cutting rules for disconnected tree diagrams following
the analysis of section 5.3 of \cite{1604.01783}. Again the proof will
proceed via induction, namely we shall prove the result assuming that it holds for diagrams with
less number of vertices. For this let us suppose that the graph contains two pieces $B$ and $C$ that
are disconnected from each other. This has been shown in Fig.~\ref{figdisone}.
$B$ and $C$ themselves may be disconnected graphs, but each
will contain less number of vertices and therefore satisfy cutting rules. Denoting by $B$ and $C$ the
expressions for the amplitudes associated with the graphs $B$ and $C$, we have the full
amplitude $A$ given by
\be 
A = -i\,  B\, C\, ,
\ee
where the factor of $-i$ arises due to the fact that the total number of disconnected components of $A$ is
equal to the sum of the number of disconnected components of $B$ and of $C$, and therefore due to
the $(i)^{1-n_c}$ factor in the expression for the amplitude with $n_c$ disconnected components, the
product $B\, C$ has one more factor of $i$ compared to $A$. This gives
\ben\label{eseven}
&& A - A^* = -i\, (B\, C+B^*\, C^*) \nonumber \\
&=& - i B^*\, C - i B\, C^* +i \, (B-B^*)\, (C-C^*) + i\, (B-B^*)\, C^* - i (B-B^*) \, C
\nonumber \\ && + i\,B^*\, (C-C^*) - i \, B\, (C-C^*) \, .
\een
The second step is the result of trivial algebraic manipulation.
The seven terms in \refb{eseven} can be diagrammatically represented as the seven
cut diagrams shown in Fig.~\ref{figdistwo}.
The additional minus
signs in the contributions from Fig~\ref{figdistwo}(c), (d) and (f), given respectively by
the third, fourth and sixth terms in \refb{eseven}, account for the fact that if $n_R$ denotes 
the number of disconnected
components to the right of the diagram, then the sum of $n_R-1$ for the component diagrams differ
from $(n_R-1)$ of the full diagram by 1.

We now note that the seven terms in 
Fig.~\ref{figdistwo} are in one to one correspondence with the cuts of $A$.
This proves the validity of the cutting rule for the disconnected amplitude $A$. 

\eject 

\noindent {\bf Acknowledgement:}

\bigskip

We would like to thank Alok Laddha and Terry Tomboulis for useful discussions.
The research of R.P.
was supported in part by Perimeter Institute for Theoretical Physics. Research at Perimeter 
Institute is supported by the Government of Canada through Industry Canada and by the Province of Ontario through the Ministry of Research and Innovation. 
This research  of A.S. was
supported in part by the 
J. C. Bose fellowship of 
the Department of Science and Technology, India.

\end{document}